\documentclass[aps,prb,reprint,amsfonts,amsmath,amssymb,longbibliography,nofootinbib,showkeys]{revtex4-2}
\usepackage{hyperref}
\usepackage{graphicx}
\usepackage{bm}
\usepackage{newtxtext}
\usepackage[varg]{newtxmath}
\usepackage[normalem]{ulem}
\usepackage{braket}
\hypersetup{colorlinks=true,linkcolor=blue,citecolor=blue,urlcolor=blue}
\usepackage{mathtools}
\usepackage[ruled,vlined]{algorithm2e}
\SetAlFnt{\footnotesize}
\SetAlCapFnt{\footnotesize}
\SetAlCapNameFnt{\footnotesize}
\usepackage{mleftright}
\mleftright
\usepackage{empheq}
\usepackage{orcidlink}

\DeclareMathOperator*{\Tr}{Tr}

\DeclareMathOperator{\re}{Re}
\DeclareMathOperator{\im}{Im}
\DeclareMathOperator{\Var}{Var}
\DeclareMathOperator{\Cov}{Cov}

\begin{document}
\title{Tangent equations of motion for nonlinear response functions}
\author{\mbox{Atsushi Ono\,\orcidlink{0000-0002-8813-4470}}}
\affiliation{\mbox{Department of Physics, Graduate School of Science, Tohoku University, Sendai 980-8578, Japan}}
\date{August 3, 2026}

\begin{abstract}
Nonlinear response functions, formulated as multipoint correlation functions or Volterra kernels, encode the dynamical and spectroscopic properties of physical systems and underpin a wide range of nonlinear transport and optical phenomena.
However, their evaluation rapidly becomes prohibitive at high orders because of combinatorial (often factorial) scaling or severe numerical errors.
Here, we establish a systematic and efficient framework to compute nonlinear response functions directly from real-time dynamics, without explicitly constructing multipoint correlators or relying on numerically unstable finite-difference methods for order-resolved extraction.
Our approach is based on the Gateaux derivative with respect to the external field in function space, which yields a closed hierarchy of tangent equations of motion (TEOM).
Propagating the TEOM alongside the original dynamics isolates each perturbative order with high accuracy, providing a term-by-term decomposition of physical contributions.
The computational cost scales exponentially with response order in the fully general setting and reduces to polynomial complexity when all perturbation directions are identical; both regimes avoid the factorial scaling of explicit multipoint-correlator evaluations.
We demonstrate the power of TEOM by computing frequency-resolved fifth-order response functions for a solid-state electron model and by obtaining nonlinear response functions up to the 49th order with controlled accuracy in a classical Duffing oscillator.
We further show that our time-evolution formulation allows optical conductivities to be evaluated directly while remaining numerically stable even near zero frequency.
TEOM can be incorporated seamlessly into existing real-time evolution methods, yielding a general framework for computing nonlinear response functions in quantum and classical dynamical systems.
\end{abstract}

\makeatletter
\renewcommand{\@keys@name}{DOI: }
\makeatother
\keywords{\href{https://doi.org/10.1103/hbjl-f5cw}{10.1103/hbjl-f5cw}}

\maketitle


\section{Introduction} \label{sec:intro}
Nonlinear responses occupy a central position in modern physics.
Linear response relates small perturbations to two-point correlation functions and provides a powerful and universal route to spectroscopy, transport, and collective dynamics \cite{Callen1951, Kubo1957, Kubo1966}, whereas higher-order responses \cite{Peterson1967, Mukamel1995, Boyd2020, Liu2022, Shen1976, Mukamel2000} encode considerably richer information:
They probe multipoint correlations, disentangle interfering excitation pathways, and reveal dynamical constraints that are invisible at the linear level.
In condensed-matter systems, nonlinear optical and dynamical responses have been utilized to access otherwise hidden collective modes, most notably the amplitude (Higgs) mode in superconductors through third-harmonic generation and related nonlinear probes \cite{Matsunaga2013, Matsunaga2014, Tsuji2015, Cea2016, Pekker2015, Shimano2020, Luo2023, Huang2025}, and to fingerprint fractionalized excitations and emergent collective modes in correlated quantum matter via multidimensional coherent spectroscopy \cite{Wan2019a, Li2021, Potts2024, Mahmood2021, Luo2023, Huang2025}.
More broadly, nonlinear responses have recently been connected to geometric aspects of quantum states and Bloch bands, stimulating renewed interest in the interplay between quantum geometry and nonlinear transport and optics \cite{Xiao2010, Sodemann2015, Kang2019, Ma2019e, Morimoto2016, DeJuan2017, Ahn2020, Ahn2022, Gao2023a, Orenstein2021, Bhalla2022, Morimoto2023}.

Beyond fundamental physics, nonlinear optical phenomena underpin a wide range of technologies, from frequency conversion and ultrafast photonics to device applications based on nonlinear materials \cite{Boyd2020, Liu2022}.
Moreover, the language of nonlinear response transcends condensed-matter physics:
It is deeply intertwined with the theory of nonlinear dynamical systems and system identification (e.g., Volterra series representations), and it provides a conceptual bridge to control engineering, where nonlinear input--output relations determine the effectiveness of manipulation protocols \cite{Schetzen1980, Khalil2001, Cheng2017}.
Across these domains, the ability to compute nonlinear response functions accurately and systematically is therefore not merely of technical interest; it is often essential for elucidating microscopic mechanisms and for designing materials and devices.

Despite this broad significance, evaluating nonlinear response functions in quantum and classical many-body systems remains a major computational challenge.
In a standard frequency-domain formulation, the $n$th-order response kernel can be written as a multipoint correlation function, whose evaluation rapidly becomes costly as the order increases.
A variety of well-established approaches exist, including sum-over-states (spectral) representations \cite{Orr1971, Boyd2020, Shen1976, Liu2022}, Green's-function and diagrammatic techniques \cite{Ward1965, Wu1988, AkbarJafari2006, Tsuji2015, Parker2019a, Holder2020, Joao2020, Tsuji2016, Tsuji2020, Silaev2019a, Chang2024, Tanabe2021, Tanaka2025, Michishita2021}, continued-fraction and polynomial-expansion methods \cite{Wan2000, Meza2019, Hallberg1995, Sota2015, Mori1965, Weise2006}, and density-matrix or superoperator formulations \cite{Mukamel1995, Sipe1993, Sipe2000, Ventura2017, Passos2018, Watanabe2021}.
These formalisms are powerful and widely used.
However, in generic many-body systems and open quantum systems, the practical bottleneck is the rapid proliferation of contributing terms with increasing order.
The combinatorial complexity of time orderings, Wick contractions, and Liouville-space pathways typically grows explosively with $n$ (often effectively factorial in naive formulations), and interactions introduce additional layers of complexity, such as vertex corrections and other many-body processes \cite{Tsuji2016, Tsuji2020, Silaev2019a, Chang2024, Tanabe2021, Tanaka2025, Kaneko2021, Kofuji2024}.
Exact diagonalization can treat interactions nonperturbatively, but finite-size effects become increasingly severe for dynamical and spectroscopic quantities in extended systems.
For dissipative dynamics, additional complications arise from non-Markovian memory and system--bath entanglement.

These difficulties have motivated the development of complementary strategies based on real-time evolution governed by an equation of motion (EOM).
Instead of constructing high-order correlators in the frequency domain, one can propagate the system under time-dependent external fields and infer response properties from the resulting time series.
This time-domain approach offers two practical advantages.
First, modern real-time solvers, such as tensor-network methods for low-dimensional correlated systems \cite{White2004, Vidal2004, Vidal2007, Haegeman2011, Holzner2011}, nonequilibrium dynamical mean-field theory for infinite-dimensional systems \cite{Aoki2014, Murakami2025}, time-dependent density functional theory for the \emph{ab initio} electronic structure of molecules and solids \cite{Runge1984, Yabana1996, Yabana2012, Marques2004, Attaccalite2011, Sato2025}, and microscopic master-equation approaches for open quantum systems \cite{Lindblad1976, Breuer2002, DeVega2017, Tanimura1989, Tanimura2020}, can incorporate interactions and dissipation within their controlled regimes while suppressing or eliminating numerical artifacts.
Second, real-time propagation treats the field-driven dynamics directly in the time domain and naturally contains contributions from all perturbative orders in the resulting time-dependent signal.
This makes it well suited for describing strongly nonlinear phenomena such as high-harmonic generation (HHG) \cite{Ghimire2011, Ghimire2019, Goulielmakis2022, Vampa2014} at the level of finite-amplitude real-time simulations.
In this sense, time evolution provides a conceptually direct route:
Given the EOM and a prescribed driving field, one can compute the full nonlinear dynamics without explicitly constructing high-order correlators or enumerating perturbative contributions order by order.

The very strength of the time-domain approach, however, also creates a central obstacle:
Because the full nonlinear dynamics contains contributions from all orders, isolating a specific perturbative order in a numerically stable and systematic manner can be highly nontrivial.
In linear response, applying a weak pulse and Fourier-transforming the induced signal is often sufficient \cite{Bertsch2000, White2004, Pereira2009, Phien2012, Shao2016}.
For nonlinear response, several time-domain protocols have been developed.
One class applies single-frequency (or nearly monochromatic) driving fields and extracts nonlinear effects from harmonic generation or from the amplitude and frequency dependence of the steady-state response (often viewed as a projection onto specific harmonic or mixing channels rather than a fully frequency-resolved kernel) \cite{Attaccalite2013, Uemoto2019, Kaneko2021, Skachkov2026, Iguchi2024, Hattori2025, Kofuji2024, Takimoto2007, Wang2007, Ding2013}.
Another class employs multiple frequency components (two-color or multicolor driving) to access frequency-mixing processes such as sum- and difference-frequency generation, and related nonlinear channels \cite{Boyd1983, Bicken2002, Guandalini2021, Pionteck2025, Kim2020}.
A third important class is multidimensional coherent spectroscopy \cite{Mukamel2000, Liu2025, Huang2026, Wan2019a, Li2021, Potts2024, Jonas2003, Cho2008, Luo2023, Huang2025}, in which multiple temporally localized pulses are applied with controlled phases and time delays, enabling highly selective access to dynamical pathways.
In these settings, techniques such as phase cycling and phase matching are used to isolate desired signal components and suppress background contributions \cite{Mukamel2000, Jonas2003, Brixner2004, Tekavec2007, Cho2008, Luttig2026}.
Across these approaches, high-order extraction can become numerically delicate, as finite-time effects and numerical errors grow with increasing order \cite{Takimoto2007, Wang2007, Ding2013, Boyd1983, Guandalini2021}.

At this point, it is important to distinguish between experimentally defined nonlinear ``signals'' and the underlying nonlinear response functions.
Multidimensional spectroscopies are commonly formulated in terms of response functions $R^{(n)}(t_n, \dots, t_1)$, but the measured signals correspond to the susceptibility tensors contracted with the field polarization vectors and then convolved with the pulse envelopes, and are further selected by phase-matching conditions and the chosen detection observable.
Consequently, ``$n$th-order signals'' in the spectroscopy literature (e.g., nominally fifth-order Raman six-wave-mixing signals, including two-dimensional Raman spectroscopy \cite{Tanimura1993, Tominaga1995, Tokmakoff1997, Blank1999, Blank2000}) represent particular projections of the underlying response and do not, in general, provide direct access to the full $n$-frequency dependence of the intrinsic $n$th-order response function $\chi^{(n)}(\omega_1,\dots,\omega_n)$.
From a computational viewpoint, order selectivity often relies on finite-amplitude subtraction, polynomial fitting, or intensity-cycling projections based on multiple real-time simulations with different field amplitudes.
While highly effective for many applications, these procedures can become increasingly expensive and ill conditioned at high orders, since lower-order components must be removed with high accuracy and numerical errors can be amplified by cancellation.

A concrete manifestation of this difficulty appears in a recent functional-derivative approach to nonlinear response extraction from time evolution.
In Ref.\ \cite{Ono2025}, we previously developed a general method in which nonlinear dynamical response functions are obtained by numerically evaluating functional derivatives of time-dependent observables with respect to the external field.
The approach is broadly applicable:
It can be combined with essentially any real-time solver, and it can, in principle, be extended beyond quantum condensed matter to more general dynamical systems.
However, the practical implementation in Ref.\ \cite{Ono2025} relied on finite-difference approximations to functional derivatives and additionally required explicit removal of lower-order contributions when extracting the $n$th-order response.
As the order increases, this combination of finite-difference step-size constraints and subtraction-induced cancellation renders the extraction increasingly ill conditioned, ultimately limiting the achievable accuracy and the accessible response order in finite-precision arithmetic.

\begin{figure}[t]\centering
\includegraphics[scale=1]{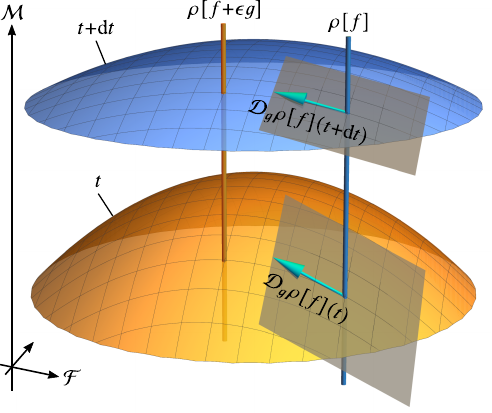}
\caption{Schematic illustration of the Gateaux derivative of the state functional $\rho[f](t)$ (e.g., a density matrix in quantum dynamics, or a state variable in classical dynamics).
The two horizontal axes represent a simplified two-dimensional slice of the function space $\mathcal{F}$ of external fields, and the vertical axis denotes the state space $\mathcal{M}$ of dynamical states.
For each time $t$, the set $\{\rho[f](t) \mid f \in \mathcal{F} \}$ forms a surface in $\mathcal{F} \times \mathcal{M}$ (orange at time $t$ and blue at time $t + \mathrm{d}t$).
The blue and yellow curves are the trajectories $\rho[f](t)$ and $\rho[f{+}\epsilon g](t)$, respectively, where $g$ is the perturbation field.
The cyan arrows on the gray tangent planes correspond to the Gateaux derivatives $\mathcal{D}_{g} \rho[f](t)$ and $\mathcal{D}_{g} \rho[f](t{+}\mathrm{d}t)$, which are tangent vectors induced by an infinitesimal variation of the external field in the direction of $g$.
The time evolution of $\mathcal{D}_{g} \rho[f](t)$ is governed by the first-order TEOM [Eq.~\eqref{eq:teom_1st}].
More generally, TEOM extend the dynamical state (quantum or classical) from $\rho[f](t)$ to a finite-order collection of field-functional derivatives, which are propagated alongside $\rho[f](t)$ and used to reconstruct the nonlinear response functions.
}
\label{fig:concept}
\end{figure}

In this work, we establish an EOM-level infinitesimal-variation framework based on the Gateaux derivative, enabling numerically stable and order-resolved extraction of nonlinear response functions from real-time dynamics.
The key idea is simple to state but far reaching in its implications:
If the system's state $\rho[f](t)$ evolves under an EOM (unitary, dissipative, or nonlinear) driven by a time-dependent external field $f(t)$, then the directional functional derivatives of $\rho[f](t)$ with respect to $f$ obey their own closed set of evolution equations (see Fig.~\ref{fig:concept} for a schematic illustration).
We refer to these as tangent equations of motion (TEOM).
The auxiliary variables are tangent vectors to the field-dependent trajectory in state space, obtained by differentiating the EOM with respect to the external field.
Throughout this paper, ``tangent'' refers to the sensitivity of the field-dependent trajectory to the infinitesimal variation $f \mapsto f + \epsilon g$ in function space, where $g$ specifies the perturbation direction and $\epsilon \to 0$; this is distinct from the usual tangent space sensitivity to initial displacements or to perturbations in finite-dimensional parameter space \cite{Benettin1980, DIMET1986, Talagrand1987, Giles2000, Ghil1991, Errico1997, Cao2003}.
Here, the perturbation is the variation of the external-field argument $f$ itself in the direction $g$; the corresponding change in observables is induced through the field dependence of the trajectory $\rho[f](t)$ and, when relevant, through the explicit field dependence of the measured operator or the dynamical generator.
This functional distinction is essential:
Differentiation with respect to $f(t)$ generates the multivariable response functions $\chi^{(n)}(\omega_1,\dots,\omega_n)$.
By deriving and integrating hierarchical TEOM for the state derivatives, we compute the $n$th-order response directly through infinitesimal variations, without any finite-difference steps and without subtracting lower-order responses.

The central advance is a reconstruction framework with two components: (1) evaluating mixed directional field-functional derivatives in the strict $\epsilon \to 0$ limit, and (2) reconstructing targeted multivariable frequency-resolved kernel slices from those derivatives using frequency- and phase-selective perturbation fields.
Component (1) is realized here by TEOM as an explicit EOM-level derivative-evaluation engine; component (2) is independent of the specific derivative engine used, as discussed further in Sec.~\ref{sec:discussion}.
This framework enables genuine frequency-resolved access beyond diagonal (harmonic) slices ($\omega_1 = \cdots = \omega_n$), including off-diagonal configurations selected by the perturbation field, while the term-by-term TEOM hierarchy also supports physically meaningful decompositions and equation-level numerical stabilization.

To highlight the scope and novelty, the central contributions of this work include:

(1) A general EOM formulation of field-functional derivatives for driven dynamics, yielding a closed TEOM hierarchy even when the dynamical generator depends on the external field and/or the evolving state, and accommodating measured operators with explicit field dependence.

(2) A frequency-resolved extraction protocol that reconstructs multivariable nonlinear response kernels $\bar{\chi}^{(n)}(\omega_1,\dots,\omega_n)$ from infinitesimal variations (Gateaux derivatives) using tailored perturbation fields, enabling off-diagonal frequency maps beyond purely harmonic (diagonal) response.

(3) A physically transparent term-by-term decomposition of nonlinear response, enabling individual contributions to be linked to microscopic processes and distinct physical mechanisms (e.g., explicit versus implicit field dependence).

(4) Demonstrations and validations across quantum and classical settings, including fifth-order frequency-resolved kernels in a two-dimensional four-band electron model and very-high-order responses (up to 49th order) in a Duffing-type oscillator.

TEOM provide a systematic hierarchy:
Alongside the original dynamics, one propagates the required mixed directional derivatives of the state up to order $n$, and lower-order derivatives are obtained simultaneously as part of the same propagation.
The hierarchy has a clear structure:
At order $n$, the number of auxiliary variables scales exponentially with $n$.
Importantly, this exponential scaling is significantly more favorable than the factorial complexity associated with enumerating large families of diagrams or evaluating explicit multipoint correlators.
Furthermore, because the differentiation is carried out directly at the EOM level, the framework applies to a broad class of field-driven dynamical systems.
In particular, it accommodates both linear and nonlinear generators of dynamics, including unitary Schr\"odinger evolution, density-matrix dynamics in Liouville space, and general nonlinear evolution equations in which the generator depends on the evolving state, as in mean-field or self-consistent dynamics.
These features make the TEOM framework systematically extendable to higher orders and enable direct interfacing with real-time solvers.

We validate the TEOM framework in a sequence of representative settings.
For noninteracting lattice electrons, we compute second- and third-order optical responses and confirm agreement with established perturbative results.
We then demonstrate the applicability to interacting dynamics in a self-consistent (mean-field) setting and compare our results qualitatively with available tensor-network results to confirm that interaction-induced spectral features are correctly captured.
Crucially, we show that the present framework provides practical access to response orders that are out of reach for existing methods such as diagrammatic techniques.
We compute fifth-order nonlinear response functions in a solid-state electron model and, as a stringent phase-sensitive benchmark, we reproduce, without parameter tuning, the polarization-angle dependence of the fifth-harmonic signal in direct comparison with independent simulations of the full driven dynamics.
We further demonstrate that very high orders can be reached in a classical nonlinear oscillator (Duffing-type) system that is also widely encountered in contexts beyond condensed-matter physics.
Along the way, the TEOM hierarchy provides equation-level stabilization and a term-by-term organization that clarifies the distinct physical contributions to the response.

The remainder of this paper is structured as follows.
In Sec.~\ref{sec:formalism}, we define the nonlinear response functions of interest and introduce the Gateaux-derivative projection scheme.
We then derive the TEOM and discuss their implementation for both linear and nonlinear generators of dynamics.
In Sec.~\ref{sec:demo}, we present numerical demonstrations and benchmarks in quantum and classical systems, including high-order response calculations.
In Sec.~\ref{sec:discussion}, we discuss the computational scaling of the TEOM-based protocol in general and in symmetric (diagonal) settings, as well as connections to alternative derivative engines, including automatic differentiation (AD).
We summarize our findings in Sec.~\ref{sec:summary}.
Technical details and complementary derivations are provided in Appendixes~\ref{sec:window}--\ref{sec:duffing_green}.

\section{Formalism} \label{sec:formalism}
In this section, we formulate an EOM approach to nonlinear response functions based on field-functional derivatives.
We first define the nonlinear response functions and their frequency-domain representations in Sec.~\ref{sec:nonlinear}.
In Sec.~\ref{sec:gateaux}, we then introduce directional functional derivatives (i.e., Gateaux derivatives) with respect to the external field and present a practical projection scheme that uses tailored perturbation fields to isolate the desired multivariable kernels from real-time signals.
Building on this, Sec.~\ref{sec:teom} derives a closed hierarchy of linearized tangent equations of motion that propagate the required mixed field derivatives of the evolving state, and discusses implementation details relevant to both linear and state-dependent (nonlinear) generators of dynamics.
In Sec.~\ref{sec:zero_field}, we discuss a zero-field source-derivative formulation as an alternative projection scheme, in which the broadband source pulse is promoted to an additional infinitesimal Gateaux-derivative direction.
Finally, in Sec.~\ref{sec:vel_gauge}, we specialize the general framework to optical responses in the velocity gauge as an instructive example in which the measured current depends explicitly on the drive, and we highlight how the time-evolution formulation avoids spurious low-frequency singularities that can arise in conventional frequency-domain transformations.

\subsection{Nonlinear response functions} \label{sec:nonlinear}
Let $\langle Q_\alpha \rangle(t)$ denote the expectation value of a physical quantity $Q_\alpha$ at time $t$.
The $n$th-order response to external fields $\{f_{\beta_i\!}(t)\}$ can be written in the general form
\begin{align}
\langle Q_\alpha \rangle^{(n)}(t)
&= \sum_{\{\beta_i\!\}} \int_{-\infty}^{\infty} \mathrm{d}t_1 \cdots \mathrm{d}t_n\, \chi_{\alpha \beta_1 \cdots \beta_n\!}^{(n)}(t; t_1, \dots, t_n) \notag \\
&\quad \times f_{\beta_1\!}(t_1) \cdots f_{\beta_n\!}(t_n),
\label{eq:Qn_t}
\end{align}
where $\chi_{\alpha \beta_1 \cdots \beta_n\!}^{(n)}(t; t_1, \dots, t_n)$ is the $n$th-order retarded response function, also known as the Volterra kernel.
The indices $\alpha$ and $\beta$ denote, for example, Cartesian directions, momentum transfer, spin quantum numbers, or combinations thereof.
Although response functions can be expressed as multipoint correlation functions, no explicit form is assumed here.

The response function in Eq.~\eqref{eq:Qn_t} has the following properties:
(1) It is nonzero only for $t \geq t_1 \geq \cdots \geq t_n$; (2) in the steady state, the time-translation invariance relation
\begin{align}
\chi^{(n)}(t; t_1, \dots, t_n) = \chi^{(n)}(t+T; t_1+T, \dots, t_n+T)
\end{align}
holds for all $T$.
This stationarity allows one to define an $n$-variable response function as
\begin{align}
\chi^{(n)}(\bar{t}_1, \dots, \bar{t}_n) = \chi^{(n)}(0; -\bar{t}_1, \dots, -\bar{t}_n),
\end{align}
with $\bar{t}_i = t - t_i$.
Then, the Fourier transform of $\langle Q_\alpha \rangle^{(n)}(t)$ is expressed as
\begin{align}
&\langle Q_{\alpha} \rangle^{(n)}(\omega) \notag \\
&= \sum_{\{\beta_i\!\}} \int_{-\infty}^{\infty} \frac{\mathrm{d}\omega_1 \cdots \mathrm{d}\omega_n}{(2\pi)^{n-1}} \delta(\omega_1 + \cdots + \omega_n - \omega) \notag \\
&\quad \times \chi_{\alpha \beta_1 \cdots \beta_n\!}^{(n)}(\omega_1, \dots, \omega_n) f_{\beta_1\!}(\omega_1) \cdots f_{\beta_n\!}(\omega_n),
\label{eq:Qn_w}
\end{align}
where $\delta$ denotes the delta function, and the response function in the frequency domain is defined by
\begin{align}
&\chi_{\alpha \beta_1 \cdots \beta_n\!}^{(n)}(\omega_1, \dots, \omega_n) \notag \\
&= \int_{-\infty}^{\infty} \mathrm{d}\bar{t}_1 \cdots \mathrm{d}\bar{t}_n\, \mathrm{e}^{\mathrm{i}\omega_1^+ \bar{t}_1 + \cdots + \mathrm{i}\omega_n^+ \bar{t}_n} \chi_{\alpha \beta_1 \cdots \beta_n\!}^{(n)}(\bar{t}_1, \dots, \bar{t}_n),
\label{eq:chin_w}
\end{align}
with $\omega_i^+ = \omega_i + \mathrm{i}0$.
When $\chi_{\alpha \beta_1 \cdots \beta_n\!}^{(n)}(\bar{t}_1, \dots, \bar{t}_n)$ is real, the following relation holds:
\begin{align}
\chi_{\alpha \beta_1 \cdots \beta_n\!}^{(n)}(\omega_{1}, \dots, \omega_{n}) = \chi_{\alpha \beta_1 \cdots \beta_n\!}^{(n)}(-\omega_{1}, \dots, -\omega_{n})^*.
\label{eq:chi_conjugate}
\end{align}

Since Eq.~\eqref{eq:Qn_w} is invariant under any permutation of index--frequency pairs, $(\beta_i, \omega_i) \leftrightarrow (\beta_j, \omega_j)$, it is useful to introduce the symmetrized $n$th-order response function
\begin{align}
&\bar{\chi}_{\alpha \beta_1 \cdots \beta_n\!}^{(n)}(\omega_1, \dots, \omega_n) \notag \\
&= \sum_{s \in \mathfrak{S}_n} \chi_{\alpha \beta_{s(1)} \cdots \beta_{s(n)}\!}^{(n)}(\omega_{s(1)}, \dots, \omega_{s(n)}),
\label{eq:symmetrized_chi}
\end{align}
where $\mathfrak{S}_n$ is the symmetric group of degree $n$.
The nonlinear response $\langle Q_\alpha \rangle^{(n)}$ can be regarded as a functional of the external fields, fully characterized by the response kernel $\bar{\chi}^{(n)}$.

\subsection{Gateaux derivative} \label{sec:gateaux}
Our objective is to extract the nonlinear response functions $\bar{\chi}^{(n)}$ from the simulated response of the physical quantity $\langle Q_\alpha \rangle(\omega)$.
To this end, we introduce the Gateaux derivative of a functional $F[f]$ at $f$, which is defined by \cite{Luenberger1997}
\begin{align}
\mathcal{D}_g F[f]
\coloneq \lim_{\epsilon \to 0} \frac{F[f + \epsilon g] - F[f]}{\epsilon}
= \left. \frac{\mathrm{d}}{\mathrm{d}\epsilon} F[f + \epsilon g] \right\vert_{\epsilon = 0},
\label{eq:def_gateaux}
\end{align}
provided that the limit exists.
Here, $g$ is a perturbation field specifying the direction of an infinitesimal variation $f \mapsto f+\epsilon g$ of the external field $f$ and has the same physical dimension as $f$; we take $\epsilon$ to be dimensionless, so that $\mathcal{D}_g F[f]$ has the same physical dimension as $F[f]$.
Although we use the Gateaux derivative for notational convenience, we assume throughout that all functionals considered in this work are $n$ times continuously Fr\'echet differentiable to the required order, which is justified within the perturbative response regime.
Under this assumption, the map $g \mapsto \mathcal{D}_g F[f]$ is linear, and more generally, the iterated derivative $\mathcal{D}_{g_n}\cdots \mathcal{D}_{g_1}F[f]$ defines a continuous symmetric $n$-linear map of $(g_1, \dots, g_n)$.

To illustrate the action of the Gateaux derivative, we first consider the second-order response
\begin{align}
\langle Q_\alpha \rangle^{(2)}(\omega)
&= \sum_{\beta_1\beta_2} \int_{-\infty}^{\infty} \frac{\mathrm{d}\omega_1 \mathrm{d}\omega_2}{2\pi} \delta(\omega_1 + \omega_2 - \omega) \notag \\
&\quad \times \chi_{\alpha \beta_1 \beta_2\!}^{(2)}(\omega_1, \omega_2) f_{\beta_1\!}(\omega_1) f_{\beta_2\!}(\omega_2).
\end{align}
The first Gateaux derivative with respect to $f_{\alpha_1}$ in the direction $g_{\alpha_1}$ is
\begin{align}
&\mathcal{D}_{g_{\alpha_1}\!} \langle Q_\alpha \rangle^{(2)}(\omega) \notag \\
&= \frac{\mathrm{d}}{\mathrm{d}\epsilon} \sum_{\beta_1\beta_2} \int_{-\infty}^{\infty} \frac{\mathrm{d}\omega_1 \mathrm{d}\omega_2}{2\pi} \delta(\omega_1 + \omega_2 - \omega) \chi_{\alpha \beta_1 \beta_2\!}^{(2)}(\omega_1, \omega_2) \notag \\
&\quad \times \bigl\{ \delta_{\alpha_1\beta_1} [f_{\beta_1\!}(\omega_1) + \epsilon g_{\beta_1\!}(\omega_1)] f_{\beta_2\!}(\omega_2) \notag \\
&\quad + \delta_{\alpha_1\beta_2} f_{\beta_1\!}(\omega_1) [f_{\beta_2\!}(\omega_2) + \epsilon g_{\beta_2\!}(\omega_2)] \bigr\} \big\vert_{\epsilon=0} \notag \\
&= \int_{-\infty}^{\infty} \mathrm{d}\omega_1\, g_{\alpha_1\!}(\omega_1) \sum_{\beta_2} \frac{\bar{\chi}_{\alpha \alpha_1 \beta_2\!}^{(2)}(\omega_1, \omega - \omega_1) f_{\beta_2\!}(\omega - \omega_1)}{2\pi},
\label{eq:Dchi2_beta}
\end{align}
where $\delta_{\alpha\beta}$ is the Kronecker delta, and
\begin{align}
&\bar{\chi}_{\alpha \alpha_1 \beta_2\!}^{(2)}(\omega_1, \omega - \omega_1) \notag \\
&= \chi_{\alpha \alpha_1 \beta_2\!}^{(2)}(\omega_1, \omega - \omega_1)
+ \chi_{\alpha \beta_2 \alpha_1\!}^{(2)}(\omega - \omega_1, \omega_1)
\label{eq:chi2_bar}
\end{align}
is the symmetrized second-order response function defined in Eq.~\eqref{eq:symmetrized_chi}.
In practical calculations, it is unnecessary to consider all components of $\{f_{\beta_i}\}$ simultaneously.
Instead, we compute the response by activating only one component, $f_{\beta_2\!} = f_{\alpha_2}$, at a time while setting all others to zero, and then repeat this procedure for each required index $\alpha_2$.
In this case, Eq.~\eqref{eq:Dchi2_beta} reduces to
\begin{align}
&\mathcal{D}_{g_{\alpha_1}\!} \langle Q_\alpha \rangle^{(2)}(\omega) \notag \\
&= \int_{-\infty}^{\infty} \mathrm{d}\omega_1\, g_{\alpha_1\!}(\omega_1) \frac{\bar{\chi}_{\alpha \alpha_1 \alpha_2\!}^{(2)}(\omega_1, \omega - \omega_1) f_{\alpha_2\!}(\omega - \omega_1)}{2\pi}.
\label{eq:DQ2}
\end{align}
Note that, throughout the paper, we do not adopt the Einstein summation convention.
The corresponding functional derivative is identified as the kernel of Eq.~\eqref{eq:DQ2}, i.e.,
\begin{align}
\frac{\delta \langle Q_\alpha \rangle^{(2)}(\omega)}{\delta f_{\alpha_1\!}(\omega_1)}
= \frac{\bar{\chi}_{\alpha \alpha_1 \alpha_2\!}^{(2)}(\omega_1, \omega - \omega_1) f_{\alpha_2\!}(\omega - \omega_1)}{2\pi}.
\end{align}
Thus, formally, the nonlinear response function can be obtained from this functional derivative.
To carry this out in practice, one may choose $g_{\alpha_1\!}(\omega)$ to approximate a delta function in the frequency domain.
Here, we consider the following two types of real-valued time-dependent fields:
\begin{align}
g_{\text{cos}_i\!}(t) &= F_0 u_{\tau_g\!}(t) \cos(\omega_i t), \label{eq:gcos_t} \\
g_{\text{sin}_i\!}(t) &= F_0 u_{\tau_g\!}(t) \sin(\omega_i t), \label{eq:gsin_t}
\end{align}
where $F_0$ is the amplitude.
The function $u_{\tau_g\!}(t)$ is an envelope function with temporal width $\tau_g$ and is chosen such that its Fourier transform satisfies
\begin{align}
u_{\tau_g\!}(\omega) = \int_{-\infty}^{\infty} \mathrm{d}t\, \mathrm{e}^{\mathrm{i}\omega t} u_{\tau_g\!}(t) \to \delta(\omega)
\label{eq:u_to_delta}
\end{align}
as $\tau_g \to \infty$.
Explicit forms of $u_{\tau_{g}\!}$ are provided in Appendix~\ref{sec:window}.
In the limit $\tau_g \to \infty$, the Fourier transforms of Eqs.~\eqref{eq:gcos_t} and \eqref{eq:gsin_t} take the forms
\begin{align}
g_{\text{cos}_i\!}(\omega) &\to \frac{F_0}{2} [ \delta(\omega - \omega_i) + \delta(\omega + \omega_i) ], \label{eq:gcos_w} \\
g_{\text{sin}_i\!}(\omega) &\to \frac{\mathrm{i} F_0}{2} [ \delta(\omega - \omega_i) - \delta(\omega + \omega_i) ], \label{eq:gsin_w}
\end{align}
and therefore, Eq.~\eqref{eq:DQ2} can be approximated by
\begin{align}
\mathcal{D}_{\text{cos}_1\!} \langle Q_\alpha \rangle^{(2)}(\omega)
&= \frac{F_0}{4\pi} \bigl[ \bar{\chi}_{\alpha \alpha_1 \alpha_2\!}^{(2)}(\omega_1, \omega-\omega_1) f_{\alpha_2\!}(\omega-\omega_1) \notag \\
&\quad + \bar{\chi}_{\alpha \alpha_1 \alpha_2\!}^{(2)}(-\omega_1, \omega+\omega_1) f_{\alpha_2\!}(\omega+\omega_1) \bigr], \\
\mathcal{D}_{\text{sin}_1\!} \langle Q_\alpha \rangle^{(2)}(\omega)
&= \frac{\mathrm{i} F_0}{4\pi} \bigl[ \bar{\chi}_{\alpha \alpha_1 \alpha_2\!}^{(2)}(\omega_1, \omega-\omega_1) f_{\alpha_2\!}(\omega-\omega_1) \notag \\
&\quad - \bar{\chi}_{\alpha \alpha_1 \alpha_2\!}^{(2)}(-\omega_1, \omega+\omega_1) f_{\alpha_2\!}(\omega+\omega_1) \bigr].
\end{align}
Here, $\mathcal{D}_{\text{cos/sin}_i\!}$ represents $\mathcal{D}_{g_{\alpha_i}\!}$, with $g_{\alpha_i\!} = g_{\text{cos/sin}_i}$.
Thus, increasing $\tau_{g}$ improves the spectral resolution along the $\omega_i$ axis (see Fig.~\ref{fig:resolution} for details).
Finally, the second-order response function is given by
\begin{align}
&\bar{\chi}_{\alpha\alpha_1\alpha_2\!}^{(2)}(\pm \omega_1, \omega \mp \omega_1) \notag \\
&= \frac{2\pi / F_0}{f_{\alpha_2\!}(\omega \mp \omega_1)} \left( \mathcal{D}_{\text{cos}_1\!} \mp \mathrm{i} \mathcal{D}_{\text{sin}_1\!} \right) \langle Q_\alpha \rangle^{(2)}(\omega).
\end{align}

The generalization to the $n$th-order response is straightforward.
The $(n-1)$th Gateaux derivative of $\langle Q_\alpha \rangle^{(n)}(\omega)$ is given by
\begin{align}
&\mathcal{D}_{g_{\alpha_{n-1}}\!} \cdots \mathcal{D}_{g_{\alpha_{1}}\!} \langle Q_\alpha \rangle^{(n)}(\omega) \notag \\
&= \int_{-\infty}^{\infty} \frac{\mathrm{d}\omega_1 \cdots \mathrm{d}\omega_{n}}{(2\pi)^{n-1}} \delta(\omega_1 + \cdots + \omega_n - \omega) \notag \\
&\quad \times \bar{\chi}_{\alpha \alpha_1 \cdots \alpha_n\!}^{(n)}(\omega_1, \dots, \omega_{n}) \notag \\
&\quad \times g_{\alpha_1\!}(\omega_1) \cdots g_{\alpha_{n-1}\!}(\omega_{n-1}) f_{\alpha_n\!}(\omega_{n}),
\label{eq:DQn}
\end{align}
from which the $n$th-order dynamical response function is formally defined as
\begin{align}
&\bar{\chi}_{\alpha \alpha_1 \cdots \alpha_n\!}^{(n)}(\omega_1, \dots, \omega_{n}) \notag \\
&= \frac{(2\pi)^{n-1}}{f_{\alpha_n\!}(\omega_{n})} \frac{\delta^{n-1} \langle Q_\alpha \rangle^{(n)}(\omega)}{\delta f_{\alpha_{n-1}\!}(\omega_{n-1}) \cdots \delta f_{\alpha_{1}\!}(\omega_{1})},
\label{eq:def_chi_nth}
\end{align}
where $\omega_n = \omega - \sum_{i=1}^{n-1} \omega_i$.
By employing the perturbation fields in Eqs.~\eqref{eq:gcos_w} and \eqref{eq:gsin_w}, the right-hand side of Eq.~\eqref{eq:def_chi_nth} can be evaluated as
\begin{empheq}[box=\fbox]{align}
&\bar{\chi}_{\alpha \alpha_1 \cdots \alpha_n\!}^{(n)}(\omega_1, \dots, \omega_{n}) \notag \\
&= \frac{(2\pi / F_0)^{n-1}}{f_{\alpha_n\!}(\omega_{n})} \mathcal{D}_{n-1}^{+} \cdots \mathcal{D}_{1}^{+} \langle Q_\alpha \rangle^{(n)}(\omega),
\label{eq:chi_nth}
\end{empheq}
where $\mathcal{D}_{i}^{\pm} = \mathcal{D}_{\text{cos}_i\!} \mp \mathrm{i} \mathcal{D}_{\text{sin}_i}$.
If $\omega_i$ is replaced by $-\omega_i$, the corresponding $\mathcal{D}_i^+$ changes to $\mathcal{D}_i^-$.
Consequently, response functions for $\omega_i < 0$ can be obtained simultaneously by sweeping only $\omega_i > 0$.
In Eqs.~\eqref{eq:DQn}--\eqref{eq:chi_nth}, only the single component of the external fields, $f_{\beta_n\!}$ with $\beta_n = \alpha_n$, is taken to be nonzero, while all other components are set to zero.

If time evolution under a complex-valued external field is possible, then by choosing the perturbation field as
\begin{gather}
g_{\text{exp}_i\!}(t) = F_0 u_{\tau_g\!}(t) \exp(-\mathrm{i}\omega_i t), \label{eq:gexp_t} \\
g_{\text{exp}_i\!}(\omega) \to F_0 \delta(\omega - \omega_i) \quad (\tau_g \to \infty),
\end{gather}
one can replace $\mathcal{D}_{i}^{\pm}$ in Eq.~\eqref{eq:chi_nth} with $\mathcal{D}_{g_{\text{exp}_i}\!}$.
For the static response, i.e., when all $\omega_i$ are zero, we have $g_{\text{sin}_i\!}(t) = 0$, and hence $\mathcal{D}_{\text{sin}_i\!} = 0$, so that $\mathcal{D}_{i}^{\pm}$ in Eq.~\eqref{eq:chi_nth} reduces to $\mathcal{D}_{\text{cos}_i}$.

Equation~\eqref{eq:chi_nth} involves derivatives of the $n$th-order response $\langle Q_{\alpha} \rangle^{(n)}(\omega)$.
In real-time simulations, however, what is directly computed is the total response $\langle Q_{\alpha} \rangle(t)$, or equivalently its Fourier transform,
\begin{align}
\langle Q_{\alpha} \rangle(\omega) = \sum_{n=0}^{\infty} \langle Q_{\alpha} \rangle^{(n)}(\omega).
\end{align}
Notably, to evaluate the Gateaux derivatives in Eq.~\eqref{eq:chi_nth}, it suffices to compute $\langle Q_{\alpha} \rangle$ at only two values of a single global field-scaling amplitude, $\pm F_0$, which uniformly scales both $f$ and all $g_i$, and then form the symmetrized and antisymmetrized combinations,
\begin{align}
\langle Q_{\alpha} \rangle^{\pm}(\omega)
= \frac{\langle Q_{\alpha} \rangle(\omega) \vert_{+F_0} \pm \langle Q_{\alpha} \rangle(\omega) \vert_{-F_0}}{2}.
\label{eq:symmetrized_Q}
\end{align}
With these definitions, $\langle Q_{\alpha} \rangle^{(n)}(\omega)$ in Eq.~\eqref{eq:chi_nth} can be replaced with $\langle Q_{\alpha} \rangle^{-}(\omega)$ for odd-order responses and $\langle Q_{\alpha} \rangle^{+}(\omega)$ for even-order responses.
This follows because the $(n-1)$th Gateaux derivative eliminates all lower-order ($< n-1$) responses, and higher-order ($> n$) contributions are negligible when $F_0$ is sufficiently small.
Since $\langle Q_{\alpha} \rangle^{\pm}(\omega)$ is either even or odd as a function of $F_0$, the next contribution beyond the $n$th order is of order $n+2$ and is further suppressed by a factor of $F_0^2$ relative to the $n$th-order term.
Furthermore, if the system preserves inversion symmetry, the parity of the response under the sign reversal of the external field is determined by the inversion parity of the observable $Q_{\alpha}$ (and the field--matter coupling).
Assuming the external field changes sign under inversion and the initial state respects inversion, if $Q_{\alpha}$ is inversion-even (inversion-odd), all odd-order (even-order) responses vanish.
In either case, only one parity sector survives; therefore, for the nonvanishing sector, the $n$th-order response $\langle Q_{\alpha} \rangle^{(n)}(\omega)$ in Eq.~\eqref{eq:chi_nth} can be replaced directly with the total response $\langle Q_{\alpha} \rangle(\omega)$.

Equation~\eqref{eq:chi_nth} recovers the finite-difference extraction scheme of Ref.\ \cite{Ono2025} for $n = 2, 3$ in the formal limit $\epsilon \to 0$.
The operational difference is that rather than approximating this limit by finite differences and subtracting nearly canceling signals, we propagate the directional derivative $\mathcal{D}_{g} \rho$ as an independent dynamical variable, as in Eq.~\eqref{eq:teom_1st} below, thereby avoiding $\epsilon$ tuning and subtractive cancellation, which become increasingly severe as the extraction order grows.

In Eqs.~\eqref{eq:gcos_w} and \eqref{eq:gsin_w}, we took the limit $\tau_g \to \infty$.
In practical simulations, however, $\tau_g$ is often taken to be finite, and it is useful to estimate its impact on the extracted response.
To roughly estimate this effect, we consider the exponential envelope in Eq.~\eqref{eq:window_exp_w} in Appendix~\ref{sec:window}.
Let $F(\omega)$ be analytic in the upper half plane and decay sufficiently fast as $\vert \omega \vert \to \infty$; then
\begin{align}
\int_{-\infty}^{\infty} \mathrm{d}\omega\, \frac{1}{\pi} \frac{\eta}{(\omega-\omega_i)^2+\eta^2} F(\omega)
= F(\omega_i + \mathrm{i}\eta),
\end{align}
with $\eta = \tau_{g}^{-1} > 0$.
Therefore, under these analyticity and convergence assumptions, the Lorentzian broadening associated with finite $\tau_g$ effectively shifts the relevant frequencies by $\mathrm{i}\eta$, thereby giving spectral peaks a finite width.
This point is discussed in detail in Sec.~\ref{sec:RM} using Figs.~\ref{fig:resolution} and \ref{fig:resolution_w1w2}.

Note that simulations in the limit $\tau_g \to \infty$, i.e., with continuous-wave (cw) perturbation fields, are also possible.
Even if $g(t) \neq 0$ at the initial simulation time, this does not affect the $n$th-order response extracted from the Gateaux derivatives of the symmetrized quantity in Eq.~\eqref{eq:symmetrized_Q}.
In practice, the main issue is convergence:
Without relaxation, a continuous wave keeps exciting the system, and the response may not converge as the simulation time increases.
This can be avoided if a dissipation term is incorporated into the EOM, as discussed in Secs.~\ref{sec:higher-order} and \ref{sec:duffing}.

\subsection{Tangent equations of motion} \label{sec:teom}
The remaining task is to evaluate the Gateaux derivatives of $\langle Q_{\alpha} \rangle(\omega)$ in Eq.~\eqref{eq:chi_nth}.
Since the Gateaux derivative and the Fourier transformation can be interchanged, we first compute the Gateaux derivatives of $\langle Q_{\alpha} \rangle(t)$ and then apply the Fourier transformation.

Let $\rho$ denote the density matrix of the system.
The expectation value of $Q_{\alpha}$ is given by
\begin{align}
\langle Q_{\alpha} \rangle(t)
= \Tr [\rho Q_{\alpha}].
\label{eq:def_expectation}
\end{align}
The density matrix evolves according to the EOM
\begin{align}
\frac{\mathrm{d}}{\mathrm{d}t} \rho = \mathcal{L} \rho,
\label{eq:eom}
\end{align}
where $\mathcal{L}$ is the Liouville operator.
In general, not only $\rho$ and $\mathcal{L}$ but also $Q_{\alpha}$ can depend on the external field, as in the case of the current operator in the velocity gauge (see Sec.~\ref{sec:vel_gauge}).

The dynamics are initiated by a pulsed external field
\begin{align}
f_{\alpha_n\!}(t) = F_0 u_{\tau_f\!}(t),
\label{eq:f_t}
\end{align}
where $F_0$ is the same amplitude as that of $g(t)$ in Eqs.~\eqref{eq:gcos_t} and \eqref{eq:gsin_t}.
In contrast to the perturbation fields $g$, the pulse width $\tau_{f}$ of the envelope function $u_{\tau_{f}\!}(t)$ is chosen to be sufficiently small, so that $f_{\alpha_n\!}(\omega_n)$ in Eq.~\eqref{eq:chi_nth} has nonzero spectral weight across the relevant energy range of the system.
Provided that this condition is satisfied, the envelope of $f$ need not be the same as that of $g$.

The first Gateaux derivative of $\langle Q_{\alpha} \rangle(t)$ is
\begin{align}
\mathcal{D}_{1} \langle Q_{\alpha} \rangle(t)
= \Tr [(\mathcal{D}_{1} \rho) Q_{\alpha}]
+ \Tr [\rho (\mathcal{D}_{1} Q_{\alpha})],
\end{align}
where $\mathcal{D}_{i}$ is a shorthand for $\mathcal{D}_{g_{\alpha_i}}$, and the second term appears when $Q_{\alpha}$ itself depends on the external fields.
The time evolution of $\mathcal{D}_{1} \rho$ in the first term is governed by
\begin{align}
\frac{\mathrm{d}}{\mathrm{d}t} \mathcal{D}_1 \rho = (\mathcal{D}_1 \mathcal{L}) \rho + \mathcal{L} (\mathcal{D}_1 \rho),
\label{eq:teom_1st}
\end{align}
with the initial condition $\mathcal{D}_1 \rho(t = -\infty) = 0$.
The Gateaux derivatives of $Q_{\alpha}$ and $\mathcal{L}$ can usually be evaluated explicitly.

We now generalize to higher-order derivatives.
Since the functionals are assumed to be sufficiently smooth, the Gateaux derivatives commute, i.e.,
\begin{align}
\mathcal{D}_i \mathcal{D}_j = \mathcal{D}_j \mathcal{D}_i.
\end{align}
We therefore introduce the notation
\begin{align}
\mathcal{D}_I = \prod_{i \in I} \mathcal{D}_{i}, \quad
\mathcal{D}_{\varnothing} = 1,
\label{eq:def_DI}
\end{align}
where $I$ denotes an index set.
The $n$th derivative of $\langle Q_{\alpha} \rangle(t)$ can then be written as
\begin{align}
\mathcal{D}_{I_n} \langle Q_{\alpha} \rangle(t) = \sum_{J \subseteq I_n} \Tr\left[ (\mathcal{D}_{J} \rho) (\mathcal{D}_{I_n \setminus J} Q_{\alpha}) \right],
\label{eq:Dn_Q}
\end{align}
where $I_n = \{1,2, \dots, n\}$, and the summation runs over all subsets $J \subseteq I_n$ (i.e., over all elements of the power set of $I_n$).
The EOM for $\mathcal{D}_{J} \rho$ in Eq.~\eqref{eq:Dn_Q} is given by
\begin{empheq}[box=\fbox]{align}
\frac{\mathrm{d}}{\mathrm{d}t} \mathcal{D}_J \rho
= \sum_{K \subseteq J} (\mathcal{D}_{K} \mathcal{L}) (\mathcal{D}_{J \setminus K} \rho),
\label{eq:Dn_rho}
\end{empheq}
with the initial condition $\mathcal{D}_{J} \rho(t = -\infty) = 0$ for $J \neq \varnothing$.
Equation~\eqref{eq:Dn_rho} recursively couples Gateaux derivatives of different orders, yielding a hierarchy of $2^n$ coupled equations for $\{\mathcal{D}_{J} \rho\}_{J \subseteq I_n}$.
Furthermore, when all perturbation directions are identical\footnote{This reduction applies whenever all Gateaux derivatives are taken along the same external-field direction.
Typical cases include static responses, where $\omega_i=0$ for all $i$, and dynamical responses to a scalar complex monochromatic perturbation, for which all derivatives are taken with respect to the same field amplitude at a common frequency $\omega_i=\omega$, as in Eq.~\eqref{eq:gexp_t} (see also Sec.~\ref{sec:duffing}).}, that is, when $\mathcal{D}_1 = \cdots = \mathcal{D}_{n}$, we have $\mathcal{D}_{J} = \mathcal{D}_1^{\vert J \vert}$, and the number of dynamical variables reduces to $\vert I_n \vert + 1 = n + 1$.
Considering that Eqs.~\eqref{eq:teom_1st} and \eqref{eq:Dn_rho} describe the time evolution of the tangent vectors $\mathcal{D}_{J} \rho$ of various orders (see Fig.~\ref{fig:concept} for a schematic illustration), we collectively refer to them as the tangent equations of motion.

The hierarchy introduced here should be distinguished from the conventional hierarchical equations of motion (HEOM) developed for non-Markovian open quantum dynamics \cite{Tanimura1989, Tanimura2020}.
In the conventional HEOM approach, the auxiliary density operators are introduced to represent the memory effects of the environment, typically through a decomposition of bath correlation functions.
In contrast, the auxiliary variables in TEOM are the field-functional derivatives $\mathcal{D}_J \rho$ of the dynamical state with respect to the external field; the hierarchy is therefore organized by response order rather than by bath-memory indices.
The two formulations are complementary rather than mutually exclusive:
If the underlying dissipative dynamics are described by HEOM, the present TEOM construction can in principle be applied by differentiating the entire HEOM hierarchy with respect to the external field.

In some applications, the Liouville operator $\mathcal{L}$ depends not only on the external field $f$ but also on the state itself, i.e., $\mathcal{L} = \mathcal{L}[f, \rho[f]]$, as in mean-field or other self-consistent dynamics.
In this case, the Gateaux derivative of the Liouville operator, $\mathcal{D}_g \mathcal{L}$, must be evaluated using the chain rule to account for its functional dependence on $\rho$.
Concretely, for a given direction $g$, one has
\begin{align}
\mathcal{D}_g \mathcal{L}
&= \left.\frac{\mathrm{d}}{\mathrm{d}\epsilon} \mathcal{L}\bigl[ f + \epsilon g, \rho[f + \epsilon g] \bigr] \right|_{\epsilon=0} \notag \\
&= (\partial_f \mathcal{L})[g] + (\partial_\rho \mathcal{L})[\mathcal{D}_g\rho], \label{eq:DL_chainrule}
\end{align}
where
\begin{align}
(\partial_f \mathcal{L})[g]
&\coloneq \left.\frac{\mathrm{d}}{\mathrm{d}\epsilon} \mathcal{L}[f + \epsilon g, \rho]\right|_{\epsilon=0}, \\
(\partial_\rho \mathcal{L})[\eta]
&\coloneq \left.\frac{\mathrm{d}}{\mathrm{d}\epsilon} \mathcal{L}[f, \rho + \epsilon \eta ]\right|_{\epsilon=0}
\end{align}
denote the Gateaux derivatives of $\mathcal{L}$ with respect to $f$ (at fixed $\rho$) and $\rho$ (at fixed $f$), respectively.
Consequently, even at first order, the TEOM acquire an additional contribution originating from the state dependence of $\mathcal{L}$, namely,
\begin{align}
\frac{\mathrm{d}}{\mathrm{d}t} \mathcal{D}_g\rho
= \bigl\{ (\partial_f\mathcal{L})[g] + (\partial_\rho\mathcal{L})[\mathcal{D}_g\rho] \bigr\}\rho
+ \mathcal{L} (\mathcal{D}_g\rho),
\end{align}
and higher-order equations involve higher (and mixed) Gateaux derivatives of $\mathcal{L}$ with respect to $\rho$.
In practice, these terms can be evaluated once the functional form of $\mathcal{L}[f, \rho]$ is specified and its functional derivatives with respect to $\rho$ are available (see Sec.~\ref{sec:MFD}).

Several remarks are in order.
First, the time evolution of the zeroth-order density matrix $\rho(t)$ follows Eq.~\eqref{eq:eom}, where the dynamics are driven solely by $f(t)$, without any contribution from the perturbation fields $g(t)$.
The influence of $g(t)$ appears only through the Gateaux derivatives.

Second, Eqs.~\eqref{eq:Dn_Q} and \eqref{eq:Dn_rho} constitute a closed set of equations that require no truncation approximations\footnote{%
``Closed'' means that, for a fixed order $n$, the dynamical state is fully specified by $\{\mathcal{D}_J\rho\}_{J\subseteq I_{n-1}}$, without introducing any additional dynamical variables.
This holds for $\mathcal{L} = \mathcal{L}[f]$, and also for $\mathcal{L} = \mathcal{L}[f,\rho]$ provided that the Gateaux derivatives of $\mathcal{L}$ with respect to $\rho$ can be expressed solely in terms of $\rho$ and $\{\mathcal{D}_J\rho\}$.
This notion of closure pertains to the TEOM hierarchy itself and is distinct from whether the underlying state equation for $\rho$ is self-contained.
For instance, if one works with a reduced state (e.g., a one-body density matrix in an interacting many-body system) whose evolution couples to higher-order reduced density matrices or correlations (a Bogoliubov--Born--Green--Kirkwood--Yvon-type hierarchy), or if non-Markovianity is represented via auxiliary variables, then one must first specify a closed dynamical model (by enlarging the state or by adopting a truncation approximation).
Once such a closed model is specified, the corresponding Gateaux derivatives can be propagated by the tangent equations without additional truncation beyond that already inherent in the chosen reduced dynamics; thereafter the tangent equations introduce no further truncation.%
}, thereby enabling accurate evaluations of the Gateaux derivatives.
A further advantage is that, when calculating the $n$th-order response function, one can simultaneously obtain the lower-order Gateaux derivatives required for the $m$th-order ($m < n$) response functions within a single simulation.

Third, the EOM [Eq.~\eqref{eq:eom}] may or may not include relaxation or dissipation terms.
If such terms are present, $\mathcal{D}_{I_n} \langle Q_{\alpha} \rangle(t)$ decays over time, determining the spectral linewidth.
If they are absent, a window function must be applied to $\mathcal{D}_{I_n} \langle Q_{\alpha} \rangle(t)$; otherwise, artifacts such as finite-size effects (recurrences) arising in long-time simulations may compromise the results.
Possible choices for the window function $w_{\tau_{w}\!}(t)$ include an exponential window, $w_{\tau_{w}\!}(t) = \exp(-\vert t \vert/\tau_w)$, a Gaussian window, $w_{\tau_{w}\!}(t) = \exp[-t^2/(2\tau_w^2)]$, or a Hann window, $w_{\tau_{w}\!}(t) = \varTheta(2\tau_{w}-\vert t \vert) \cos^2[\pi t/(4\tau_{w})]$, where $\varTheta$ is the unit step function (see also Appendix~\ref{sec:window}).
Here, $\tau_w$ denotes the temporal width of the window function, and a larger $\tau_w$ yields higher spectral resolution for the response frequency $\omega = \sum_{i=1}^{n} \omega_i$.

Finally, we emphasize that the key concept of the TEOM formalism is broadly applicable.
Equation~\eqref{eq:def_expectation} gives the expectation value expressed in terms of the density matrix $\rho$ and its Gateaux derivatives.
Alternatively, one can formulate the theory in terms of the state vector $\vert \psi \rangle$ that evolves according to the Schr\"odinger equation,
\begin{align}
\frac{\mathrm{d}}{\mathrm{d}t} \vert \psi \rangle
= - \frac{\mathrm{i}}{\hbar} \mathcal{H} \vert \psi \rangle,
\label{eq:eom_schroedinger}
\end{align}
where $\mathcal{H}$ is the Hamiltonian, and then calculate the Gateaux derivatives using the product rule:
\begin{gather}
\mathcal{D}_J \langle Q_{\alpha} \rangle
= \sum_{A \sqcup B \sqcup C = J}
\langle \mathcal{D}_{A} \psi \vert (\mathcal{D}_{B}Q_{\alpha}) \vert \mathcal{D}_{C}\psi \rangle,
\\
\frac{\mathrm{d}}{\mathrm{d}t} \vert \mathcal{D}_J \psi \rangle
= - \frac{\mathrm{i}}{\hbar} \sum_{K \subseteq J} (\mathcal{D}_K \mathcal{H}) \vert \mathcal{D}_{J\setminus K} \psi \rangle.
\end{gather}
Here, the first summation runs over all ordered set partitions of $J$ into (possibly empty) disjoint subsets $A,B,C$, and we assume that the Gateaux derivatives are taken with respect to real perturbation fields so that $\mathcal{D}_J \langle \psi \vert = \langle \mathcal{D}_J \psi \vert$.
This wave-function formalism may be particularly suitable for tensor-network and exact-diagonalization methods.
As discussed above, the present framework can also be extended to cases where the Liouville operator in Eq.~\eqref{eq:eom} depends on $\rho$, leading to nonlinear dynamics, such as mean-field dynamics (MFD) in interacting many-body systems (Sec.~\ref{sec:MFD}) and an anharmonic oscillator (Sec.~\ref{sec:duffing}).
It further applies to systems governed by the time-dependent Ginzburg--Landau equation for the superconducting order parameter or by the Bloch equation describing Anderson pseudospins in superconductors \cite{Tsuji2015} and excitonic insulators \cite{Kaneko2021}.
Furthermore, it encompasses EOM such as the Landau--Lifshitz--Gilbert equation for classical spins and the Euler--Lagrange, Hamilton, Liouville, or Boltzmann equations for classical dynamical systems (Sec.~\ref{sec:duffing}).
Therefore, the present framework can be applied to any dynamical system, whether quantum or classical, provided that a closed and sufficiently smooth EOM is specified and the resulting TEOM can be integrated numerically within the regime of validity of the chosen model.

For practical implementation, the workflow is summarized in Algorithm~\ref{alg:teom_time}.
The subsequent Fourier-domain reconstruction and (if needed) cosine/sine projection that realize Eq.~\eqref{eq:chi_nth} are summarized in Algorithm~\ref{alg:teom_recon}.
A compact Julia implementation of this workflow for the spinless Rice--Mele model discussed in Sec.~\ref{sec:RM} is publicly available on Zenodo \cite{ZenodoTEOMRiceMele}.

\begin{algorithm}[t]
\caption{Time-domain TEOM propagation.}
\label{alg:teom_time}
\DontPrintSemicolon
\SetAlgoLined

\KwIn{Order $n$ ($m=n-1$, $I_m=\{1,\dots,m\}$); time grid $\{t_k\}$ with step $\delta t$;
EOM $\dot\rho=\mathcal{L}\rho$ (possibly nonlinear in $f$ and $\rho$);
observable $Q_\alpha$ (possibly field dependent);
pulse $f(t)=F_0 u_{\tau_f\!}(t)$;
mode: (A) real with fixed $s\in\{\mathrm{cos},\mathrm{sin}\}^m$ or (B) complex;
perturbation fields $g_i(t)$ given by the chosen mode [Eqs.~\eqref{eq:gcos_t}, \eqref{eq:gsin_t}, \eqref{eq:gexp_t}].}
\KwOut{Time series $y(t)$:
(A) $y(t)=\mathcal{D}_{s_m}\cdots\mathcal{D}_{s_1}\langle Q_\alpha\rangle(t)$,\;
(B) $y(t)=\mathcal{D}_m^{+}\cdots\mathcal{D}_1^{+}\langle Q_\alpha\rangle(t)$.}

\BlankLine
\textbf{State tensors:} store $\rho_J(t)\equiv \mathcal{D}_J\rho(t)$ for all $J\subseteq I_m$
(and $Q_J\equiv \mathcal{D}_J Q_\alpha$ if $Q_\alpha$ is field dependent; otherwise $Q_J=0$ for $J\neq\varnothing$).\;
\tcp{Implementation: represent $J\subseteq I_m$ by a bitmask and store $\{\rho_J\}$ in an array of length $2^m$.}
Initialize at time $t = t_0$: $\rho_\varnothing(t_0)=\rho(t_0)$ and $\rho_J(t_0)=0$ for $J\neq\varnothing$.\;
Precompute or provide routines to evaluate $\mathcal{L}_K(t)\equiv (\mathcal{D}_K\mathcal{L})(t)$ for all $K\subseteq I_m$
[using the chain rule in Eq.~\eqref{eq:DL_chainrule} if needed].\;

\BlankLine
\ForEach{time step $t\to t+\delta t$}{
  Evaluate $f(t)$ and $\{g_i(t)\}_{i\in I_m}$.\;
  Compute the $m$-th derivative of the expectation value:
  $\displaystyle y(t) = \sum_{J\subseteq I_m}\Tr[\rho_{J}(t) Q_{I_m\setminus J}(t)]
  \qquad \text{[Eq.~\eqref{eq:Dn_Q}]}.$\;
  Store $y(t)$.\;
  Evaluate $\{\mathcal{L}_K(t)\}_{K\subseteq I_m}$ at the current fields (and state, if applicable).\;
  Integrate all $\{\rho_J\}_{J\subseteq I_m}$ simultaneously using the same time-stepping scheme:
  $\displaystyle \frac{\mathrm{d}}{\mathrm{d}t} \rho_{J}(t) = \sum_{K \subseteq J} \mathcal{L}_{K}(t) \rho_{J \setminus K}(t)
  \qquad \text{[Eq.~\eqref{eq:Dn_rho}]}.$\;
}

Repeat for all desired grid points of $(\omega_1,\dots,\omega_m)$.\;
In mode (A), also repeat for all $s\in\{\mathrm{cos},\mathrm{sin}\}^m$; in mode (B), no $s$-loop.\;
All runs are independent and parallelizable.\;
\end{algorithm}

\begin{algorithm}[t]
\caption{Frequency-domain reconstruction of $\bar{\chi}^{(n)}$.}
\label{alg:teom_recon}
\DontPrintSemicolon
\SetAlgoLined

\KwIn{Fixed $(\omega_1,\dots,\omega_m)$ with $m=n-1$.
Mode (A): time series $\{y_{s}(t)\}_{s\in\{\mathrm{cos},\mathrm{sin}\}^m}$,
or mode (B): complex time series $y^{+\!}(t)$.
Optional $\pm F_0$ pairs; window $w(t)$; pulse spectrum $f_{\alpha_n\!}(\omega_n)$.}

\KwOut{$\bar{\chi}^{(n)}_{\alpha\alpha_1\cdots\alpha_n\!}(\sigma_1\omega_1,\dots,\sigma_m\omega_m,\omega_n)$
as a function of $\omega$ for $\sigma\in\Sigma$, where
$\Sigma=\{\pm1\}^m$ in (A) and $\Sigma=\{(+1,\dots,+1)\}$ in (B).}

\BlankLine
\textbf{(Optional symmetrization)}\;
\uIf{both $\pm F_0$ simulations are available}{
  \tcp{Apply this to each available time series: all $y_s(t)$ in (A), or $y^+(t)$ in (B).}
  Form $y_\bullet^{\pm\!}(t) = \bigl(y_\bullet|_{+F_0}\pm y_\bullet|_{-F_0}\bigr)/2$ [Eq.~\eqref{eq:symmetrized_Q}].\;
  Set $y_\bullet^{(n)\!}(t) = y_\bullet^{-\!}(t)$ for odd $n$ and $y_\bullet^{(n)\!}(t)=y_\bullet^{+\!}(t)$ for even $n$.\;
}\Else{
  Set $y_\bullet^{(n)\!}(t) = y_\bullet(t)$.\;
}

\BlankLine
\textbf{Fourier transforms}\;
\uIf{mode (A)}{
  \ForEach{$s\in\{\mathrm{cos},\mathrm{sin}\}^m$}{
    $Y_s(\omega) = \int \mathrm{d}t\, \mathrm{e}^{\mathrm{i}\omega t} w(t) y_s^{(n)\!}(t)$.\;
  }
}\Else(\tcp*[f]{mode (B)}){%
  $Y_+(\omega) = \int \mathrm{d}t\, \mathrm{e}^{\mathrm{i}\omega t} w(t) y^{+(n)\!}(t)$.\;
}

\BlankLine
\textbf{Projection and reconstruction}\;
\ForEach{$\sigma=(\sigma_1,\dots,\sigma_m)\in\Sigma$}{
  \uIf{mode (A)}{
    \tcp{Basis change: $\mathcal{D}_i^{\sigma_i} = \mathcal{D}_{\mathrm{cos}_i} - \mathrm{i} \sigma_i \mathcal{D}_{\mathrm{sin}_i}$}
    Define $C_{\sigma,s}=\prod_{i=1}^m c_i$ with
    $c_i = 1$ if $s_i = \mathrm{cos}$ and $c_i = - \mathrm{i} \sigma_i$ if $s_i = \mathrm{sin}$.\;
    Set $Y_\sigma(\omega)=\sum_{s\in\{\mathrm{cos},\mathrm{sin}\}^m} C_{\sigma,s}\, Y_s(\omega)$.\;
  }\Else{
    Set $Y_\sigma(\omega)=Y_+(\omega)$ [only $\sigma=(+1,\dots,+1)$].\;
  }

  Set $\omega_n(\sigma)=\omega-\sum_{i=1}^m \sigma_i\omega_i$.\;
  $\displaystyle \bar{\chi}^{(n)}_{\alpha\alpha_1\cdots\alpha_n\!}(\sigma_1\omega_1,\dots,\sigma_m\omega_m,\omega_n(\sigma))
  = \frac{(2\pi/F_0)^{n-1}}{f_{\alpha_n\!}(\omega_n(\sigma))} Y_\sigma(\omega)
  \qquad \text{[Eq.~\eqref{eq:chi_nth}]}.$\;
}
\end{algorithm}

\subsection{Zero-field source-derivative formulation} \label{sec:zero_field}

In this subsection, we introduce an alternative formulation in which the broadband source pulse is treated as an infinitesimal Gateaux-derivative direction, rather than as a finite weak field.
The formulation introduced in Secs.~\ref{sec:gateaux} and \ref{sec:teom}, and used in the numerical demonstrations below, keeps the source field $f(t)$ finite and extracts the target order by the symmetrization or antisymmetrization in Eq.~\eqref{eq:symmetrized_Q}.
Although this parity projection is controlled and is not a finite-difference approximation to the Gateaux derivative, it can still involve cancellations between finite-amplitude signals. 
Moreover, because the projection is performed at finite $F_0$, residual higher-order contributions remain, and convergence with respect to the weak-source amplitude $F_0$ has to be checked.

These issues can be removed by taking one additional Gateaux derivative along the direction of the broadband source pulse.
We introduce
\begin{align}
s_{\alpha_n\!}(t) = F_0 u_{\tau_s\!}(t),
\end{align}
which has the same temporal profile as the weak source field $f(t)$ in Eq.~\eqref{eq:f_t}.
We use the symbol $s$ to emphasize that this field is now a Gateaux-derivative direction, while the reference external field $f$ is set to zero after differentiation.

Taking the additional derivative of Eq.~\eqref{eq:DQn} with respect to the source direction gives
\begin{align}
&\mathcal{D}_{s_{\alpha_n}\!} \mathcal{D}_{g_{\alpha_{n-1}}\!} \cdots \mathcal{D}_{g_{\alpha_1}\!} \langle Q_\alpha \rangle^{(n)}(\omega)
\notag \\
&= \int_{-\infty}^{\infty} \frac{\mathrm{d}\omega_1 \cdots \mathrm{d}\omega_n}{(2\pi)^{n-1}} \delta(\omega_1+\cdots+\omega_n-\omega) \notag \\
&\quad \times \bar{\chi}^{(n)}_{\alpha\alpha_1\cdots\alpha_n\!}(\omega_1,\ldots,\omega_n)
g_{\alpha_1\!}(\omega_1)\cdots g_{\alpha_{n-1}\!}(\omega_{n-1}) s_{\alpha_n\!}(\omega_n).
\end{align}
The integration over $\omega_n$ can be performed using the delta function, yielding
\begin{align}
&\mathcal{D}_{s_{\alpha_n}\!} \mathcal{D}_{g_{\alpha_{n-1}}\!} \cdots \mathcal{D}_{g_{\alpha_1}\!} \langle Q_\alpha \rangle^{(n)}(\omega) \notag \\
&= \int_{-\infty}^{\infty} \frac{\mathrm{d}\omega_1 \cdots \mathrm{d}\omega_{n-1}}{(2\pi)^{n-1}} \bar{\chi}^{(n)}_{\alpha\alpha_1\cdots\alpha_n\!}(\omega_1,\ldots,\omega_n) \notag \\
&\quad \times g_{\alpha_1\!}(\omega_1)\cdots g_{\alpha_{n-1}\!}(\omega_{n-1}) s_{\alpha_n\!}(\omega_n),
\end{align}
where $\omega_n = \omega - \sum_{i=1}^{n-1}\omega_i$.
Therefore, the reconstruction formula corresponding to Eq.~\eqref{eq:chi_nth} becomes
\begin{align}
&\bar{\chi}^{(n)}_{\alpha\alpha_1\cdots\alpha_n\!}(\omega_1,\ldots,\omega_n)
\notag \\
&= \frac{(2\pi/F_0)^{n-1}}{s_{\alpha_n\!}(\omega_n)}
\mathcal{D}_{s_{\alpha_n}\!} \mathcal{D}_{n-1}^{+} \cdots \mathcal{D}_{1}^{+} \langle Q_\alpha \rangle(\omega) \big\vert_{f=0}.
\label{eq:chi_nth_zero_field}
\end{align}
Here, the derivative is applied to the total response $\langle Q_\alpha \rangle$, not to its $n$th-order component $\langle Q_\alpha \rangle^{(n)}$.
This is because the $n$ Gateaux derivatives annihilate all terms of order lower than $n$, whereas any term of order higher than $n$ retains at least one undifferentiated factor of the reference external field and therefore vanishes when evaluated at $f=0$.
Thus, the target $n$th-order response is isolated without the finite-amplitude symmetrization in Eq.~\eqref{eq:symmetrized_Q}.

The TEOM hierarchy itself is unchanged.
In the weak-source formulation, the derivative-direction set has size $m = n-1$, whereas in the zero-field source-derivative formulation, this set is enlarged to $m=n$ by adding the source direction $s_{\alpha_n}$.
In this sense, the present formulation treats all perturbative external-field legs as tangent directions.
It is therefore the cleaner formulation from a formal point of view:
The finite physical source field is removed from the extraction procedure, the symmetrization or antisymmetrization in Eq.~\eqref{eq:symmetrized_Q} is no longer required, and no perturbative convergence check with respect to the finite source amplitude is needed; $F_0$ only fixes the normalization of the Gateaux-derivative direction.

Furthermore, in the equilibrium applications considered in this work, the zeroth-order trajectory in this formulation is the field-free trajectory.
Since the initial state is chosen to be a stationary state of the field-free EOM, one has $\dot{\rho}=0$ and hence $\rho(t)$ is constant.
Thus, the zeroth-order state does not require an additional real-time propagation; only the tangent variables associated with the Gateaux-derivative directions have to be propagated.
For a nonequilibrium reference state driven by a strong pump field, in contrast, the zeroth-order reference trajectory would obey the pump-driven EOM and would have to be propagated in real time.

This formal advantage, however, comes with an increased computational cost.
If $m$ denotes the number of independent Gateaux-derivative directions propagated in a single TEOM hierarchy, the zero-field source-derivative formulation has $m=n$, whereas the weak-source formulation used in the numerical calculations has $m=n-1$.
The number of auxiliary derivative variables scales as $2^m$.
Moreover, the direct TEOM right-hand side contains the subset couplings in Eq.~\eqref{eq:Dn_rho}; hence, a generic direct implementation has an upper-bound scaling
\begin{align}
\sum_{J\subseteq I_m}2^{|J|}=3^m,
\label{eq:subset_couplings}
\end{align}
where $I_m$ is the set of the $m$ derivative directions.
Thus, promoting the broadband source pulse to an additional Gateaux-derivative direction increases not only the memory footprint but also the cost of evaluating the TEOM right-hand side.
In a simple implementation for the Rice--Mele model studied in Sec.~\ref{sec:RM}, the zero-field source-derivative formulation increased the computation time by factors of approximately $2.30$, $2.57$, and $2.81$ for the second-, third-, and fourth-order calculations, respectively.
These factors are model and implementation dependent, but they illustrate the additional cost associated with the extra Gateaux-derivative direction.

For this reason, the numerical demonstrations in Sec.~\ref{sec:demo} use the weak-source formulation.
The zero-field source-derivative formulation provides the formally cleaner projection, while the weak-source formulation provides a reduced-hierarchy realization that is advantageous in practice when the finite-source convergence can be controlled.

\subsection{Example: optical responses in the velocity gauge} \label{sec:vel_gauge}
Building on the general framework presented above, we examine the optical responses in the velocity gauge.
This example is particularly instructive because it provides a paradigmatic case in which the physical quantity (the electric current) couples nonlinearly to the external field (the vector potential) and depends explicitly on it.
While the length-gauge formalism has been widely adopted for nonlinear optical response calculations owing to its manifest gauge invariance and robustness near zero frequency \cite{Aversa1995, Sipe2000, Taghizadeh2017, Ventura2017}, the velocity gauge has also been employed, especially in lattice models for strongly correlated systems.
A key advantage of the velocity gauge is that it preserves spatial translation symmetry, which facilitates real-time simulations.

We consider the Hamiltonian
\begin{align}
\mathcal{H} = \sum_{\bm{k}mn} c_{\bm{k}m}^\dagger H_{mn}(\bm{k}) c_{\bm{k}n},
\label{eq:hamiltonian_general}
\end{align}
where $c_{\bm{k}n}$ is the annihilation operator for an electron with wave vector $\bm{k}$ and quantum number $n$.
In the velocity gauge, the light--matter coupling is introduced through the Peierls substitution with a spatially uniform vector potential $\bm{A}(t)$, i.e.,
\begin{align}
\bm{k} \to \bm{k} - \frac{e}{\hbar} \bm{A}(t),
\end{align}
where $e$ denotes the electric charge ($e < 0$ for electrons).
Since the matrix $H(\bm{k})$ in Eq.~\eqref{eq:hamiltonian_general} depends nonlinearly on $\bm{k}$, the electric current operator
\begin{align}
\bm{\mathcal{J}}[\bm{A}] = - \frac{\partial \mathcal{H}[\bm{A}]}{\partial \bm{A}}
\label{eq:def_current}
\end{align}
also depends on the vector potential.
The expectation value of the electric current density is defined by
\begin{align}
\langle \bm{j} \rangle(t)
= \frac{1}{N V} \sum_{\bm{k}} \Tr[\rho \bm{J}]
\equiv \frac{1}{N V} \sum_{\bm{k}} \sum_{mn} \rho_{nm} \bm{J}_{mn},
\label{eq:def_current_density}
\end{align}
where $\rho_{nm}(\bm{k}) = \langle c_{\bm{k}m}^\dagger c_{\bm{k}n} \rangle$ is the one-body density matrix, $\bm{J} = (e/\hbar) \partial H(\bm{k}-e\bm{A}/\hbar)/\partial \bm{k}$ is the matrix of the current operator, $N$ is the number of $\bm{k}$ points, and $V$ is the volume of the unit cell.
The time evolution of the one-body density matrix follows the von Neumann equation,
\begin{align}
\frac{\mathrm{d}}{\mathrm{d}t} \rho = \mathcal{L} \rho = -\frac{\mathrm{i}}{\hbar} [H, \rho],
\label{eq:vonNeumann}
\end{align}
where both $H$ and $\rho$ depend on the external field.

The Gateaux derivative of the current density is given by
\begin{align}
\mathcal{D}_1 \langle \bm{j} \rangle(t)
= \frac{1}{N V} \sum_{\bm{k}} \left\{ \Tr[(\mathcal{D}_1 \rho) \bm{J}] + \Tr[\rho (\mathcal{D}_1 \bm{J})] \right\},
\label{eq:current_decomposition}
\end{align}
and $\mathcal{D}_1 \rho$ in the first term follows the TEOM,
\begin{align}
\frac{\mathrm{d}}{\mathrm{d}t} \mathcal{D}_1 \rho
= -\frac{\mathrm{i}}{\hbar} [\mathcal{D}_1 H, \rho]
-\frac{\mathrm{i}}{\hbar} [H, \mathcal{D}_1 \rho].
\end{align}
The Gateaux derivatives of the Hamiltonian and current-operator matrices are expressed as
\begin{gather}
\mathcal{D}_{1} H
= - \frac{e}{\hbar} g_{\alpha_1} \frac{\partial H(\bm{k} - e\bm{A}/\hbar)}{\partial k_{\alpha_1}}, \label{eq:D1_H} \\
\mathcal{D}_{1} J_\alpha
= \frac{e}{\hbar} \left(- \frac{e}{\hbar} g_{\alpha_1} \right) \frac{\partial^2 H(\bm{k} - e\bm{A}/\hbar)}{\partial k_{\alpha_1} \partial k_{\alpha}},
\label{eq:D1_J}
\end{gather}
where we treat the vector potential as the external field, i.e., $\bm{A}(t) = \bm{f}(t)$.
Higher-order derivatives can also be obtained explicitly.

In conventional frequency-domain approaches, such as diagrammatic methods, the vector-potential dependence of the electric current operator complicates the calculation of response functions.
Even at first order, it is necessary to consider not only the linear response of the paramagnetic current, which is the zeroth-order term in $\bm{A}$ in Eq.~\eqref{eq:def_current}, but also the equilibrium (unperturbed) contribution from the diamagnetic current, which is already linear in $\bm{A}$.
The situation becomes increasingly complex for higher-order responses.

On the other hand, approaches based on real-time simulations directly evaluate the $n$th-order response function
\begin{align}
&\bar{\chi}_{\alpha \alpha_1 \cdots \alpha_n\!}^{(n)}(\omega_1, \dots, \omega_{n}) \notag \\
&= \frac{(2\pi)^{n-1}}{A_{\alpha_n\!}(\omega_{n})} \frac{\delta^{n-1} \langle j_\alpha \rangle^{(n)}(\omega)}{\delta A_{\alpha_{n-1}\!}(\omega_{n-1}) \cdots \delta A_{\alpha_{1}\!}(\omega_{1})},
\label{eq:def_optical_chi}
\end{align}
without computing the various types of multipoint correlation functions required in frequency-domain formalisms.
We adopt the velocity gauge and fix the remaining gauge freedom by choosing $\bm{A}(t) \to 0$ as $t \to -\infty$, so that
\begin{align}
\bm{A}(t) = - \int_{-\infty}^{t} \mathrm{d}t'\, \bm{E}(t'),
\label{eq:A_and_E}
\end{align}
where $\bm{E}(t)$ is the electric field.
With this convention, the nonlinear optical conductivity
\begin{align}
&\bar{\sigma}_{\alpha \alpha_1 \cdots \alpha_n\!}^{(n)}(\omega_1, \dots, \omega_{n}) \notag \\
&= \frac{(2\pi)^{n-1}}{E_{\alpha_n\!}(\omega_{n})} \frac{\delta^{n-1} \langle j_\alpha \rangle^{(n)}(\omega)}{\delta E_{\alpha_{n-1}\!}(\omega_{n-1}) \cdots \delta E_{\alpha_{1}\!}(\omega_{1})}
\label{eq:def_optical_sig}
\end{align}
can also be obtained by identifying $f$ and $g$ with $\bm{E}$ and applying $\bm{A}$ given by Eq.~\eqref{eq:A_and_E} in the actual calculation.
In conventional approaches, one first computes $\chi^{(n)}$ and then uses the relation $\bm{E}(\omega) = \mathrm{i} \omega \bm{A}(\omega)$ to derive
\begin{align}
\bar{\sigma}_{\alpha \alpha_1 \cdots \alpha_n\!}^{(n)}(\omega_1, \dots, \omega_{n})
= \frac{\bar{\chi}_{\alpha \alpha_1 \cdots \alpha_n\!}^{(n)}(\omega_1, \dots, \omega_{n})}{\mathrm{i}^n \omega_1 \cdots \omega_n}.
\label{eq:sig_and_chi}
\end{align}
A practical difficulty of this frequency-domain conversion is that factors $1/\omega_i$ make it undefined or numerically unstable whenever $\omega_i \to 0$.
The time-evolution approaches avoid this explicit low-frequency conversion:
Computing $\bar{\sigma}^{(n)}$ directly via Eq.~\eqref{eq:def_optical_sig} yields a quantity that remains well defined even at $\omega_i = 0$ and eliminates the spurious oscillations that arise from the denominator, as demonstrated quantitatively in Fig.~\ref{fig:rm_sig}.

It should be emphasized that the TEOM construction itself is not tied to the velocity gauge.
If a closed and differentiable length-gauge EOM is specified, the same Gateaux-differentiation procedure can be applied.
In extended periodic systems, the length gauge requires the usual care associated with the position operator, Berry connections, and covariant derivatives in the Bloch representation.
These issues are inherited from the underlying length-gauge formulation rather than introduced by TEOM.
Once a gauge-covariant length-gauge EOM is adopted, the corresponding TEOM follow by differentiating it with respect to the external field.

\section{Numerical demonstration} \label{sec:demo}
In this section, we numerically validate the formalism developed in Sec.~\ref{sec:formalism} and demonstrate its usefulness through several computational examples.
In Sec.~\ref{sec:RM}, we compute the nonlinear optical response functions of the simplest lattice model for ferroelectrics and examine the convergence of our calculations.
We also present numerical results for the nonlinear optical conductivities, following the framework discussed in Sec.~\ref{sec:vel_gauge}.

In Sec.~\ref{sec:MFD}, we treat a setting in which the Liouville operator depends on the density matrix itself, as occurs in MFD.
As a concrete example, we calculate a nonlinear response function for an interacting many-body system and show that it agrees qualitatively with tensor-network results.
Furthermore, by decomposing the TEOM contributions to $\mathcal{D}_1\langle j\rangle$ [Eq.~\eqref{eq:current_decomposition}], we identify which term is responsible for the long-lived dc-like component and its strong window-width (relaxation-time) sensitivity, thereby providing a practical diagnostic of potentially problematic (approximation-induced) contributions without rejecting MFD wholesale.

In Sec.~\ref{sec:higher-order}, we investigate a two-dimensional four-band system that describes electrons coupled to a topological spin texture and show that our framework enables calculations of frequency-resolved response functions up to the fifth order.
Using these results, we explain the polarization-angle dependence of high-harmonic generation in the perturbative regime.

In Sec.~\ref{sec:duffing}, we consider the classical Duffing oscillator as a distinct example.
It has far fewer degrees of freedom (dynamical variables) than quantum systems, allowing us to compute response functions to much higher orders.
We also obtain the exact response functions through the Green's-function method and use them as an exact reference to validate the accuracy of our method.
Finally, we discuss how the computational cost scales with the response order $n$.

\subsection{Nonlinear optical responses in the Rice--Mele model} \label{sec:RM}
We consider the spinless Rice--Mele model \cite{Rice1982}, which is a minimal model for ferroelectrics.
The Hamiltonian matrix in Eq.~\eqref{eq:hamiltonian_general} is given by
\begin{align}
H(k)
= -h_x \cos(ka/2)\, \tau_x - h_y \sin(ka/2)\, \tau_y - h_z \tau_z,
\end{align}
where $\tau_x$, $\tau_y$, and $\tau_z$ are the Pauli matrices describing the sublattice degree of freedom, and $a$ denotes the lattice constant.
The parameters $h_x$ and $h_y$ represent the transfer integrals, with $h_y$ quantifying the strength of the bond alternation, while $h_z$ determines the on-site potential difference between the two sublattices.
When both $h_y/h_x$ and $h_z/h_x$ are nonzero, inversion symmetry is broken, and even-order optical responses are allowed.
The eigenvalues of $H(k)$ are given by
\begin{align}
\varepsilon_{k}^{\pm}
= \pm \sqrt{h_x^2 \cos^2(ka/2) + h_y^2 \sin^2(ka/2) + h_z^2},
\end{align}
and the linear optical excitation spectrum exhibits a continuum in the range $\hbar \omega \in [2\varepsilon_{k=\pi/a}^+, 2\varepsilon_{k=0}^+]$ when $\vert h_y \vert < \vert h_x \vert$.

We introduce the vector potential $\bm{A}(t)$ through the Peierls substitution and discuss the response of the electric current $\langle j \rangle = (Na)^{-1} \sum_{k} \Tr[\rho J]$, with $J = (e/\hbar) \partial H(k-eA/\hbar)/\partial k$ the current-operator matrix.
The time evolution of the one-body density matrix $\rho(t)$ is governed by Eq.~\eqref{eq:vonNeumann}, with the initial state $\rho(t = -\infty)$ corresponding to a half-filled insulating system.

Hereafter, we set $h_x = 1$ as the unit of energy.
With this choice, all physical quantities are expressed in units constructed from $e$, $\hbar$, $a$, and $h_x$, which are all set to unity.
For example, the units of time, vector potential, electric field, and electric current are given by $\hbar/h_x$, $\hbar/(ea)$, $h_x/(ea)$, and $eh_x/\hbar$, respectively.

Unless otherwise stated, the parameters are set to $h_y = 0.5$, $h_z = 0.1$, $N = 800$, $F_0 = 10^{-6}$, $\tau_f = 0.1$, and $\tau_g = \tau_w = 100$.
For these parameters, the one-photon excitation spectrum has a continuum in the range $1.02 \lesssim \omega \lesssim 2.01$ (see also Fig.~2 of Ref.\ \cite{Ono2025}).
The time evolution, governed by the von Neumann equation \eqref{eq:vonNeumann}, is simulated from $t = -500$ to $1000$ using the fourth-order Runge--Kutta method.
The time step is $\delta t = 2^{-7}$ for $\lvert t \rvert < 6\tau_f$ and $\delta t = 2^{-4}$ otherwise.

\begin{figure}[t]\centering
\includegraphics[scale=1]{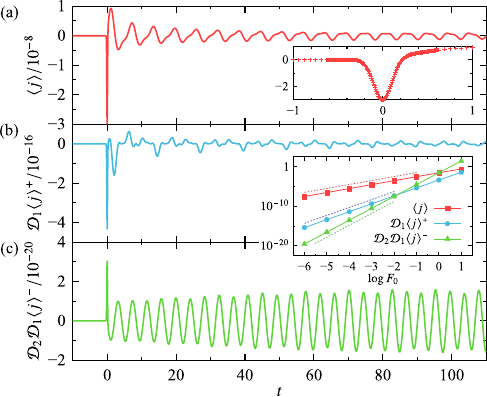}
\caption{Temporal profiles of the electric current and its Gateaux derivatives with $g_{\alpha_1\!} = g_{\text{cos}_1\!}$ and $\omega_1 = 1.5$.
(a) Electric current.
The inset shows an enlarged view for $\vert t \vert < 1$, where the crosses indicate the data points calculated on the adaptive time grid.
(b) First Gateaux derivative of the symmetrized electric current.
(c) Second Gateaux derivative of the antisymmetrized electric current.
The inset spanning panels (b) and (c) presents the absolute values of the currents at $t = 0$ as functions of the field amplitude, and the dashed lines indicate power-law scalings $\propto F_0$ (red), $\propto F_0^2$ (blue), and $\propto F_0^3$ (green).
}
\label{fig:rm_cur}
\end{figure}

In Fig.~\ref{fig:rm_cur}, we show the temporal profiles of the electric current and its Gateaux derivatives.
Here, we use a cosine-type perturbation field with a Gaussian envelope and central frequency $\omega_1 = \omega_2 = 1.5$.
When a Gaussian vector potential $A(t) = f(t)$ centered at $t = 0$ is applied, oscillations are induced in the current and its Gateaux derivatives.
Both $\langle j \rangle$ and $\mathcal{D}_1 \langle j \rangle^+$ exhibit damped oscillations, which are attributed to interference among electrons excited into the energy continuum.
Although no damping is observed for $\mathcal{D}_2 \mathcal{D}_1 \langle j \rangle^-$ up to $t \sim 110$, damping does appear on longer timescales.
Note that $\langle j \rangle^+$ and $\langle j \rangle^-$ are the symmetrized and antisymmetrized currents, respectively, as defined by Eq.~\eqref{eq:symmetrized_Q}.
Before (anti)symmetrization, the $(n-1)$th Gateaux derivative of the current has a leading term of order $F_0^{n-1}$ and exhibits a time dependence that follows the broad envelope of $g(t)$.

The inset spanning panels (b) and (c) of Fig.~\ref{fig:rm_cur} plots the values of the current and its Gateaux derivatives at $t = 0$ as functions of the field amplitude $F_0$.
As expected, they exhibit the power-law scaling $\langle j \rangle \propto F_0$, $\mathcal{D}_1 \langle j \rangle^+ \propto F_0^2$, and $\mathcal{D}_2 \mathcal{D}_1 \langle j \rangle^- \propto F_0^3$, indicating that the Gateaux derivatives systematically isolate the contribution at each order.

\begin{figure}[t]\centering
\includegraphics[scale=1]{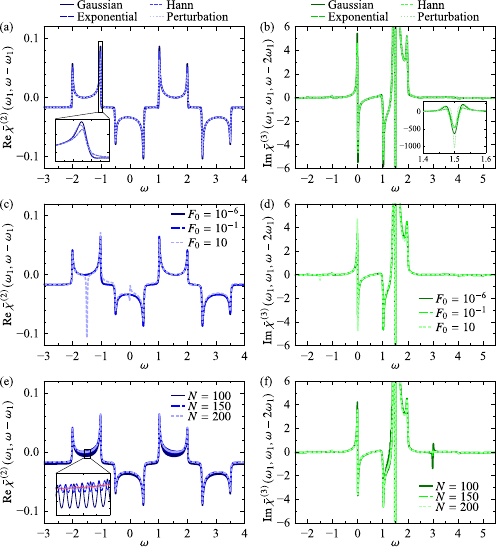}
\caption{Convergence behavior of the second- and third-order response functions, for $\omega_1 = 1.5$.
(a), (b) Response functions obtained with Gaussian, exponential, and Hann windows, compared with perturbation-theory results.
The inset in panel (a) is an enlarged view from $\omega = -1.1$ to $-0.95$.
(c), (d) Dependence on the field amplitude.
(e), (f) Dependence on the number of $k$ points.
The inset in panel (e) shows an enlarged view from $\omega = -1.6$ to $-1.4$, where the red curve corresponds to $N = 800$.
}
\label{fig:rm_check}
\end{figure}

By applying a Fourier transform to the temporal profiles, we can evaluate the response functions through Eq.~\eqref{eq:chi_nth}.
Figure~\ref{fig:rm_check} shows the second- and third-order response functions, $\bar{\chi}^{(2)}(\omega_1, \omega-\omega_1)$ (left panels) and $\bar{\chi}^{(3)}(\omega_1, \omega_2, \omega-\omega_1-\omega_2)$ (right panels), as functions of $\omega$, for $\omega_1 = \omega_2 = 1.5$.

The top panels of Fig.~\ref{fig:rm_check} compare the results obtained with different window functions for the Fourier transform, namely, the Gaussian, exponential, and Hann windows (see Appendix~\ref{sec:window}), together with the results from conventional perturbation theory.
Overall, the perturbation-theory results are reproduced irrespective of the window choice, indicating the validity of our framework.
The differences among window functions become pronounced near the band edges, where Van Hove singularities occur.
Among the three, the exponential window shows the closest agreement with perturbation theory.
The Gaussian window provides higher resolution of the band-edge structure, and the Hann window offers intermediate resolution between the Gaussian and exponential windows.
These window functions should be chosen according to the purpose:
For example, the Gaussian or Hann window is preferable when higher resolution is needed, whereas the exponential window is better suited for quantitative comparisons with perturbation theory.
If relaxation or dissipation processes are included in the EOM, one should choose the window width $\tau_w$ (and the envelope width $\tau_g$ of the perturbation fields) such that the resulting Fourier resolution is sufficient to resolve the spectral linewidth determined by the lifetime of excitations (see also Secs.~\ref{sec:higher-order} and \ref{sec:duffing} for examples in the $\tau_g \to \infty$ and $\tau_w \to \infty$ limits).

The middle panels of Fig.~\ref{fig:rm_check} show the response functions obtained at several values of the field amplitude $F_0$.
The results for $F_0 = 10^{-1}$ and $10^{-6}$ are nearly indistinguishable.
In contrast, a pronounced deviation appears for $F_0 = 10$, because contributions of order $n+2$ and higher contaminate the estimate of $\bar{\chi}^{(n)}$.
When evaluating the response functions through Eq.~\eqref{eq:chi_nth}, $F_0 = 10^{-6}$ can be regarded as sufficiently small and is an appropriate choice for the present system.
In practice, we must verify that the extracted $\bar{\chi}^{(n)}$ is independent of $F_0$; lack of convergence may indicate either numerical errors or a breakdown of the perturbative expansion in the external-field amplitude.

The bottom panels of Fig.~\ref{fig:rm_check} present the dependence on the number of $k$ points, $N$.
For $N = 100$, artificial oscillations due to finite-size effects appear in the excitation continua.
When $N$ is increased to $200$, the spectra become nearly smooth.
The enlarged view (inset) of Fig.~\ref{fig:rm_check} includes the result for $N = 800$ shown as a red curve, indicating that $N = 800$ is sufficient even at this level of magnification.
Note that the dependence on the number of $k$ points (or lattice sites) can change substantially depending on the required spectral resolution, that is, on the temporal widths $\tau_g$ and $\tau_w$.
In general, achieving higher spectral resolution requires a larger $N$.

\begin{figure}[t]\centering
\includegraphics[scale=1]{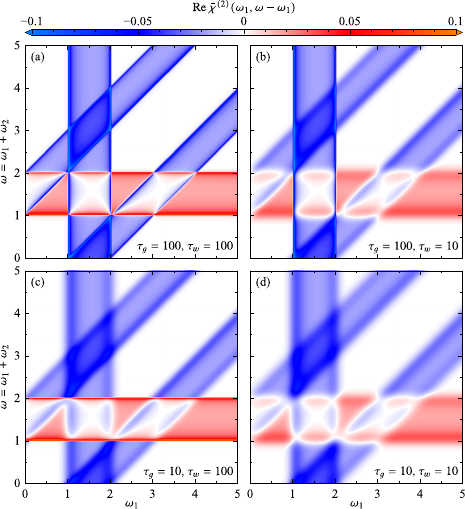}
\caption{Dependence of spectral resolution on the temporal widths of the Gaussian perturbation field $\tau_{g}$ and the Gaussian window function $\tau_{w}$, for (a) $\tau_{g} = \tau_{w} = 100$, (b) $\tau_{g} = 100$ and $\tau_{w} = 10$, (c) $\tau_{g} = 10$ and $\tau_{w} = 100$, and (d) $\tau_{g} = \tau_{w} = 10$.
}
\label{fig:resolution}
\end{figure}

Having established convergence with respect to the window choice, field amplitude, and $k$ point sampling, we next examine how the perturbation-field width $\tau_g$ and the window-function width $\tau_w$ affect the spectral resolution of the response functions.
Figure~\ref{fig:resolution} shows the second-order response function $\bar{\chi}^{(2)}(\omega_1, \omega-\omega_1)$ obtained by sweeping the central frequency $\omega_1$ of the perturbation field.
Here, $\omega = \omega_1 + \omega_2$ is the response frequency, and both the perturbation field $g(t)$ and the window function $w(t)$ are taken to be Gaussian waveforms.
Taking the case $\tau_g = \tau_w = 100$ in Fig.~\ref{fig:resolution} as a reference, Fig.~\ref{fig:resolution}(b), with $\tau_w = 10$, shows reduced resolution along the $\omega$ axis, while preserving the resolution along the $\omega_1$ axis.
Conversely, when $\tau_g = 10$ and $\tau_w = 100$ [Fig.~\ref{fig:resolution}(c)], the resolution along the $\omega_1$ axis is reduced.
This behavior is consistent with the respective roles of the envelope and window functions.

\begin{figure}[t]\centering
\includegraphics[scale=1]{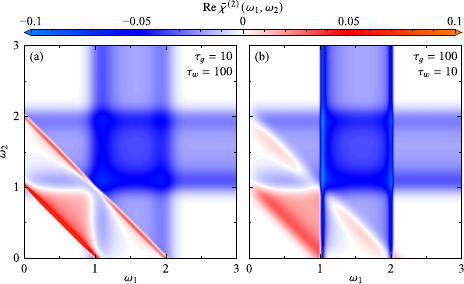}
\caption{Same data as Fig.~\ref{fig:resolution}, but replotted as a function of $(\omega_1, \omega_2)$.
(a) $\tau_{g} = 10$ and $\tau_{w} = 100$; (b) $\tau_{g} = 100$ and $\tau_{w} = 10$.
}
\label{fig:resolution_w1w2}
\end{figure}

Figure~\ref{fig:resolution_w1w2} shows the same data as Fig.~\ref{fig:resolution}, replotted as a function of $(\omega_1, \omega_2)$.
In Fig.~\ref{fig:resolution_w1w2}(a), where $\tau_g = 10$ and $\tau_w = 100$, the resolutions along the $\omega_1$ and $\omega_2$ directions are comparable, whereas the resolution along the diagonal direction is higher, scaling as $\tau_{w}^{-1}$.
In contrast, in Fig.~\ref{fig:resolution_w1w2}(b), where $\tau_g = 100$ and $\tau_w = 10$, only $\omega_1$ exhibits high resolution on the order of $\tau_{g}^{-1}$, while the resolutions along the $\omega_2$ and $\omega$ directions appear to scale as $\tau_{w}^{-1}$.

This anisotropic spectral resolution can be understood by examining how the reconstruction treats the input frequencies and the response frequency.
The $n$th-order response function obtained from Eq.~\eqref{eq:chi_nth} is naturally expressed as a function of $(\omega_1,\omega_2,\dots,\omega_{n-1},\omega)$, where the last variable is the response frequency $\omega = \sum_{i=1}^{n} \omega_i$.
For Gaussian envelope and window functions, the spectral resolution along each input-frequency axis $\omega_1,\dots,\omega_{n-1}$ is controlled by the perturbation-field width and scales as $\Delta\omega_i \sim \tau_g^{-1}$, whereas the resolution along the response-frequency axis $\omega$ scales as $\Delta\omega \sim \tau_w^{-1}$.
It is convenient to interpret these finite resolutions as effective uncertainties (standard deviations) on the frequency axes\footnote{``Effective uncertainties'' are understood as the second-moment width of the resolution kernel, not as statistical noise variances.
For non-Gaussian kernels, we may instead characterize resolution by a robust width measure; the scalings discussed here should be understood at the level of order of magnitude.}.
Namely, we regard $\omega_1,\dots,\omega_{n-1}$ and $\omega$ as ``measured'' variables with uncertainties $\sqrt{\Var(\omega_i)} \sim \tau_g^{-1}$ and $\sqrt{\Var(\omega)} \sim \tau_w^{-1}$, respectively.
Since $\tau_g$ and $\tau_w$ can be tuned independently and act on different axes, as shown in Fig.~\ref{fig:resolution}, we neglect correlations between the uncertainties of $\omega$ and $\omega_i$, and we assume\footnote{This assumption is made in the $(\omega_1, \dots, \omega_{n-1}, \omega)$ representation; after transforming to $(\omega_1, \dots, \omega_n)$, nonzero covariances generally appear.}
\begin{equation}
\Cov(\omega_i, \omega) \approx 0, \quad
\Cov(\omega_i, \omega_j) \approx 0,
\end{equation}
for $1 \leq j < i < n$.
Under this approximation, the effective resolution of the last input frequency $\omega_n = \omega - \sum_{i=1}^{n-1}\omega_i$ follows from standard error propagation:
\begin{equation}
\Var(\omega_n)
\approx
\Var(\omega) + \sum_{i=1}^{n-1} \Var(\omega_i)
\sim
\tau_w^{-2} + (n-1) \tau_g^{-2}.
\label{eq:omega_n_resolution}
\end{equation}
Equation~\eqref{eq:omega_n_resolution} implies that the resolution associated with $\omega_n$ is generally worse than that of $\omega_1,\dots,\omega_{n-1}$, reflecting the fact that $\omega_n$ is inferred as a linear combination of the ``measured'' frequencies $(\omega_1, \dots, \omega_{n-1}, \omega)$.
While increasing $\tau_w$ sharpens the resolution along the response-frequency axis $\omega$, the resolution kernel becomes increasingly anisotropic when expressed in the $(\omega_1, \dots, \omega_n)$ space.
Specifically, the width along the $\omega$ direction shrinks as $\Delta\omega \sim \tau_w^{-1}$, whereas the width along each $\omega_i$ direction remains of order $\Delta\omega_i \sim \tau_g^{-1}$, as shown in Fig.~\ref{fig:resolution_w1w2}(a).
For this reason, we should avoid taking $\tau_w$ much larger than $\tau_g$, which would over-resolve the $\omega$ direction and lead to a highly direction-dependent resolution kernel in $(\omega_1, \dots, \omega_n)$.
A pragmatic choice is to keep $\tau_w$ on the same order as $\tau_g$; in the results below, we set $\tau_w = \tau_g$ for simplicity.
If the goal is a quantitative comparison with conventional methods (e.g., Green's-function approaches), however, the priority is to incorporate into the EOM a dissipative term that is comparable to that used in those methods; in that case, the best agreement is typically obtained by taking the continuous-wave limit $\tau_g \to \infty$ and using no window function, i.e., $\tau_w \to \infty$ (see also Secs.~\ref{sec:higher-order} and \ref{sec:duffing}).

\begin{figure*}[t]\centering
\includegraphics[scale=1]{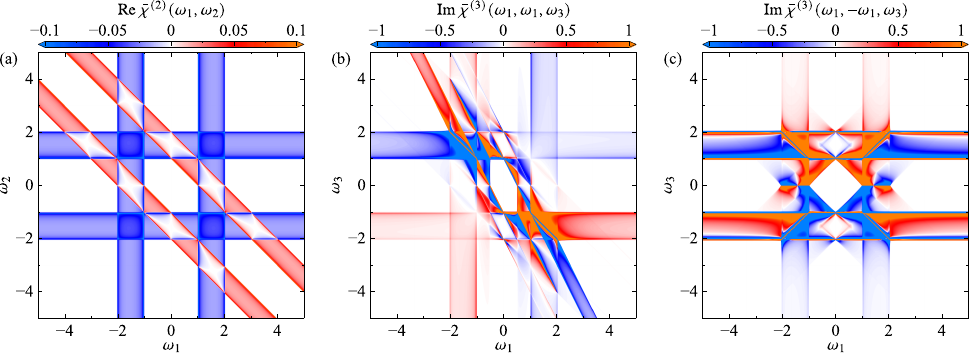}
\caption{Color maps of the second- and third-order response functions in the Rice--Mele model.
(a) Real part of the second-order response function $\bar{\chi}^{(2)}(\omega_1, \omega_2)$.
Imaginary parts of the third-order response functions (b) $\bar{\chi}^{(3)}(\omega_1, \omega_1, \omega_3)$ and (c) $\bar{\chi}^{(3)}(\omega_1, -\omega_1, \omega_3)$.
}
\label{fig:rm_chi}
\end{figure*}

In Fig.~\ref{fig:rm_chi}, we show the second- and third-order nonlinear response functions, computed with the Gaussian perturbation field and the Gaussian window function with $\tau_g = \tau_w = 100$.
The second-order response function, $\bar{\chi}^{(2)}(\omega_1,\omega_2)$, is obtained over the two-dimensional frequency space $(\omega_1, \omega_2)$ by scanning only $\omega_1$ (with $\omega_1 \geq 0$).
For the third-order response function, we scan $\omega_1$ while setting $\omega_2 = \omega_1$, and plot $\bar{\chi}^{(3)}(\omega_1,\omega_1,\omega_3)$ and $\bar{\chi}^{(3)}(\omega_1,-\omega_1,\omega_3)$ in the $(\omega_1, \omega_3)$ plane.
These results reproduce the perturbation-theory results and satisfy the relation in Eq.~\eqref{eq:chi_conjugate}.
Note that the present computation achieves higher accuracy and greater numerical stability than that in Ref.\ \cite{Ono2025}, because our TEOM framework eliminates cancellation errors arising from the finite-difference (finite-$\epsilon$) approximation and the lower-order subtraction.

For $\re \bar{\chi}^{(2)}(\omega_1, \omega_2)$ [Fig.~\ref{fig:rm_chi}(a)], we observe two distinct features: negative continua along the vertical and horizontal directions for $1 \lesssim \vert \omega_{i} \vert \lesssim 2$ ($i = 1,2$), and positive continua along the diagonal direction for $1 \lesssim \vert \omega_1 + \omega_2 \vert \lesssim 2$.
The former is attributed to a one-photon resonance, whereas the latter arises from a two-photon resonance.
In regions where these continua overlap, destructive interference suppresses the magnitude of $\bar{\chi}^{(2)}$.

For $\im \bar{\chi}^{(3)}(\omega_1, \omega_2, \omega_3)$ with $\omega_2 = \omega_1$, shown in Fig.~\ref{fig:rm_chi}(b), we observe continua in the same frequency ranges as the resonance features seen in Fig.~\ref{fig:rm_chi}(a).
In addition, a vertical continuum corresponding to a two-photon resonance emerges for $1 \lesssim \vert 2\omega_1 \vert \lesssim 2$, and a continuum associated with a three-photon resonance appears for $1 \lesssim \vert 2\omega_1 + \omega_3 \vert \lesssim 2$.
In the regions where these continua overlap, $\bar{\chi}^{(3)}$ exhibits an intricate pattern of sign changes.

The behavior of $\im \bar{\chi}^{(3)}(\omega_1, \omega_2, \omega_3)$ with $\omega_2 = -\omega_1$, shown in Fig.~\ref{fig:rm_chi}(c), can be understood in the same way.
Specifically, in addition to the continua present in $\bar{\chi}^{(2)}$, diagonal continua appear for $1 \lesssim \vert \omega_3 \pm \omega_1 \vert \lesssim 2$, corresponding to a two-photon resonance.

\begin{figure}[t]\centering
\includegraphics[scale=1]{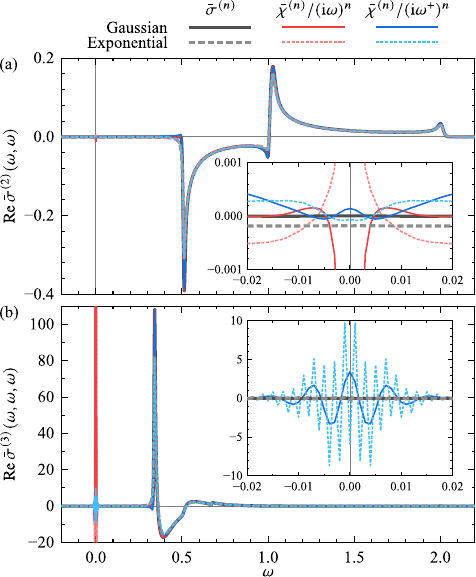}
\caption{Nonlinear optical conductivities in the Rice--Mele model.
(a) Second-order optical conductivity and (b) third-order optical conductivity, calculated using the Gaussian window (solid, darker colors) and the exponential window (dashed, lighter colors), with $\tau_{w} = 100$.
The gray lines are obtained by calculating $\bar{\sigma}^{(n)}$ directly from Eq.~\eqref{eq:def_optical_sig}, whereas the red and blue lines are obtained from $\bar{\chi}^{(n)}$ using Eq.~\eqref{eq:sig_and_chi}.
The insets show a magnified view for $\vert \omega \vert < 0.02$.
In the inset of panel (b), the results for $\bar{\chi}^{(3)}/(\mathrm{i}\omega)^3$ (red lines) are omitted because numerical artifacts would obscure the other curves.
}
\label{fig:rm_sig}
\end{figure}

Finally, we present the results for the nonlinear optical conductivity defined in Eq.~\eqref{eq:def_optical_sig}.
Figure~\ref{fig:rm_sig} plots the real parts of $\bar{\sigma}^{(2)}(\omega,\omega)$ and $\bar{\sigma}^{(3)}(\omega,\omega,\omega)$, together with the corresponding quantities obtained from the current-response function $\bar{\chi}^{(n)}$ via Eq.~\eqref{eq:sig_and_chi}.

When evaluating $\bar{\sigma}^{(n)}$ from $\bar{\chi}^{(n)}$, we consider two procedures.
In the first, we keep $\omega_i$ in the denominator of Eq.~\eqref{eq:sig_and_chi} unchanged, yielding $\bar{\chi}^{(n)}/(\mathrm{i}\omega)^n$, which is undefined at $\omega_i = 0$.
In the second, we introduce a small imaginary part $\eta = \tau_{w}^{-1}$ \cite{Passos2018}, also called a broadening factor, through $\omega_i \to \omega_i + \mathrm{i}\eta$, yielding $\bar{\chi}^{(n)}/(\mathrm{i}\omega^+)^n$.
These two results are shown as the red and blue lines, respectively.

Although all methods give qualitatively similar behavior overall, an important difference appears near $\omega = 0$.
Without the broadening factor, $\bar{\chi}^{(n)}/(\mathrm{i}\omega)^n$ diverges at $\omega = 0$, and it exhibits strong spurious oscillations in the vicinity of $\omega = 0$.
With the broadening factor, $\bar{\chi}^{(n)}/(\mathrm{i}\omega^+)^n$ remains well defined at $\omega = 0$; however, as shown in the insets of Fig.~\ref{fig:rm_sig}, its behavior near $\omega = 0$ still contains unphysical structures.

In contrast, direct evaluation of $\bar{\sigma}^{(n)}$ from Eq.~\eqref{eq:def_optical_sig} remains well defined at $\omega = 0$ and almost completely suppresses the artificial oscillations near $\omega = 0$, yielding a physically reasonable low-frequency behavior.
With the Gaussian window function, both $\bar{\sigma}^{(2)}$ and $\bar{\sigma}^{(3)}$ become nearly zero at $\omega = 0$, consistent with the insulating nature of the ground state.
For the exponential window, however, $\bar{\sigma}^{(2)}$ retains a small but finite value at $\omega = 0$, as seen in the inset of Fig.~\ref{fig:rm_sig}(a).
Since the exponential window mimics a phenomenological damping with rate $\eta = \tau_{w}^{-1}$, this residual value can be interpreted as a damping- or broadening-induced contribution rather than an intrinsic dc response of the insulating ground state \cite{Wilhelm2021, Murakami2022}.

These observations highlight an important distinction between directly evaluating the electric-field response and converting a vector-potential response through Eq.~\eqref{eq:sig_and_chi}.
The latter procedure introduces explicit low-frequency conversion factors, $1/(\omega_1\cdots\omega_n)$, and can therefore amplify small residual errors in $\bar{\chi}^{(n)}$ near $\omega_i = 0$ into unphysical structures in $\bar{\sigma}^{(n)}$.
This point is especially relevant when a window function or phenomenological relaxation rate is used as an effective broadening.
In the velocity gauge, physical relaxation or decoherence terms must in general be introduced in a gauge-consistent manner and may acquire an explicit dependence on the vector potential~\cite{Tokman2009, Wismer2018}.
If this gauge consistency is not enforced, residual artifacts in $\bar{\chi}^{(n)}$ may be further enhanced by the low-frequency conversion factors.
In contrast, Eq.~\eqref{eq:def_optical_sig} treats the electric field itself as the perturbing field and therefore avoids this explicit division by powers of frequency.
Field-dependent dissipative terms, when needed, can be incorporated in the TEOM framework through the corresponding Gateaux derivatives of the Liouvillian.

\subsection{Many-body effects through mean-field dynamics} \label{sec:MFD}
To discuss many-body effects in the nonlinear response, we consider the Rice--Mele--Hubbard model, which extends the Rice--Mele model studied in Sec.~\ref{sec:RM} by including spin and an on-site Hubbard interaction.
The Hamiltonian is defined as
\begin{align}
\mathcal{H}
&= -\sum_{js} \frac{h_x - (-1)^j h_y}{2} \left( c_{j+1,s}^\dagger c_{js} + \text{H.c.} \right) \notag \\
&\quad - h_z \sum_{js} (-1)^j n_{js}
+ U \sum_{j} n_{j\uparrow} n_{j\downarrow},
\end{align}
where $c_{js}^\dagger$ creates an electron with spin $s$ at site $j$, $n_{js} = c_{js}^\dagger c_{js}$ is the number operator, and $U$ denotes the strength of the Hubbard interaction.
The parameters $(h_x, h_y, h_z)$ are the same as those in the spinless Rice--Mele model.

Instead of treating the Hubbard interaction exactly using a tensor-network method as in Ref.\ \cite{Ono2025}, we employ the Hartree--Fock approximation and consider the time evolution of mean fields, hereafter referred to as MFD.
This approach can incorporate many-body effects associated with vertex corrections, as discussed in Refs.\ \cite{Kaneko2021, Tanabe2021, Kofuji2024, Tanaka2025}.
We perform a standard Hartree--Fock decoupling of the interaction term $U n_{j\uparrow} n_{j\downarrow}$, retaining the Fock (spin off-diagonal) contributions in addition to the Hartree terms to preserve spin-rotational symmetry.
We also assume a two-sublattice structure.
Under these approximations and assumptions, the Hartree--Fock Hamiltonian can be written as
\begin{align}
\mathcal{H}_{\text{HF}}
= \sum_{kmn}  c_{km}^\dagger H_{mn}(k) c_{kn},
\end{align}
where
\begin{align}
H(k) = H_0(k) + H_{\mathrm{int}},
\end{align}
with the noninteracting part
\begin{align}
H_0(k) &= -h_x \cos\left(\frac{ka}{2}\right)\, \sigma_0 \otimes \tau_x
- h_y \sin\left(\frac{ka}{2}\right)\, \sigma_0 \otimes \tau_y \notag \\
&\quad - h_z\, \sigma_0 \otimes \tau_z,
\end{align}
and the interacting part
\begin{align}
H_{\mathrm{int}} = \frac{U}{2} \sum_{\tau = \uparrow,\downarrow} \bigl[ \langle \sigma_0 \otimes \tau_{\tau} \rangle_{\rho} \sigma_0 \otimes \tau_{\tau} - \langle \bm{\sigma} \otimes \tau_{\tau} \rangle_{\rho} \cdot \bm{\sigma} \otimes \tau_{\tau} \bigr].
\end{align}
Here, $\sigma_\alpha$ and $\tau_\alpha$ are the Pauli matrices (with the identity matrices for $\alpha = 0$) acting on the spin and sublattice degrees of freedom, respectively, and $\tau_{\uparrow/\downarrow} = (\tau_{0} \pm \tau_{z})/2$.
The expectation value is defined as
\begin{align}
\langle \bullet \rangle_{\rho} = \frac{1}{Na} \sum_{k} \Tr[\rho\, \bullet],
\label{eq:rmh_expectationvalue}
\end{align}
where $N$ ($= 800$) is the number of $k$ points.
The lattice constant $a$ is chosen to be the length of a two-site unit cell; that is, the nearest-neighbor distance is $a/2$.

We adopt the velocity gauge, introducing the vector potential through the Peierls substitution $k \to k - e A(t)/\hbar$ with $A(t) = f(t)$.
The expectation value of the electric current is defined by Eq.~\eqref{eq:def_current_density}; however, to reproduce the results for the spinless Rice--Mele model in the noninteracting limit $U = 0$, we multiply it by an additional factor of $1/2$.

The Hamiltonian matrix $H(k)$ contains terms that depend on the expectation value $\langle \bm{\sigma} \otimes \tau_{\uparrow/\downarrow} \rangle_{\rho}$, indicating that $H(k)$ is a functional of both $A$ and $\rho$.
Therefore, the first Gateaux derivative of $H(k)$ is given by
\begin{align}
\mathcal{D}_1 H
= (\partial_f H)[g_1] + (\partial_\rho H)[\mathcal{D}_1 \rho],
\end{align}
where
\begin{align}
(\partial_f H)[g_1] = -\frac{e}{\hbar} g_{1} \frac{\partial H(k - eA/\hbar)}{\partial k}
\end{align}
is the one-dimensional case of Eq.~\eqref{eq:D1_H}, and
\begin{align}
&(\partial_\rho H)[\mathcal{D}_1 \rho] \notag \\
&= \frac{U}{2} \sum_{\tau = \uparrow,\downarrow} \bigl[ \langle \sigma_0 \otimes \tau_{\tau} \rangle_{\mathcal{D}_1 \rho} \sigma_0 \otimes \tau_{\tau} - \langle \bm{\sigma} \otimes \tau_{\tau} \rangle_{\mathcal{D}_1 \rho} \cdot \bm{\sigma} \otimes \tau_{\tau} \bigr]
\end{align}
provides an additional contribution that arises from the MFD.

If one is interested in third- and higher-order responses, higher-order Gateaux derivatives of $H$ are required, and in general this includes mixed derivatives with respect to $f$ and $\rho$.
However, for models in which the Hamiltonian can be decomposed as
\begin{align}
H[f, \rho] = H_{0}[f] + H_{\mathrm{int}}[\rho],
\end{align}
all mixed Gateaux derivatives vanish, e.g., $\partial_{f} \partial_{\rho} H = 0$, which simplifies the higher-order analysis.

As in Sec.~\ref{sec:RM}, we set $h_x = 1$ as the energy unit and choose $h_y = 0.5$ and $h_z = 0.1$.
We consider $U$ in the range $U \leq 0.2$ and restrict our analysis to the half-filled system.
For these parameters, the ground state is a paramagnetic ferroelectric insulator.
We simulate the time evolution using the fourth-order Runge--Kutta method with a time step of $\delta t = 2^{-7}$ for $\lvert t \rvert < 6\tau_f$ and $\delta t = 2^{-4}$ otherwise.

\begin{figure}[t]\centering
\includegraphics[scale=1]{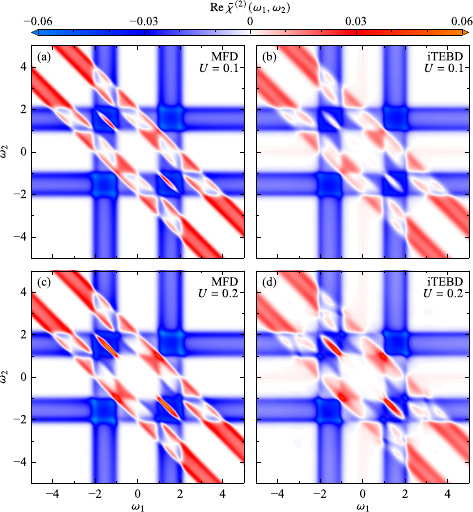}
\caption{Real part of the second-order response function in the Rice--Mele--Hubbard model, for $U = 0.1$ (upper panels) and $0.2$ (lower panels).
MFD are included in the left panels, whereas the iTEBD method is employed in the right panels (adapted from Ref.\ \cite{Ono2025}).
The temporal widths of the Gaussian envelope function and the Gaussian window functions are set to $\tau_g = \tau_w = 10$ in the MFD results, and to $\tau_g = 12$ and $\tau_w = 6$ in the iTEBD results.
All results are symmetrized in postprocessing to be consistent with Eq.~\eqref{eq:symmetrized_chi}.
}
\label{fig:rmh_map}
\end{figure}

Figure~\ref{fig:rmh_map} shows $\re \bar{\chi}^{(2)}(\omega_1, \omega_2)$ for $U = 0.1$ and $0.2$.
The left panels present the results including MFD, whereas the right panels show the results obtained by infinite time-evolving block decimation (iTEBD) \cite{Vidal2007} reported in Ref.\ \cite{Ono2025}, which are based on the finite-difference approximation.
We do not aim for pointwise quantitative agreement here:
MFD and iTEBD operate at different approximation levels and use different effective broadenings, $\tau_g$ and $\tau_w$.
The purpose of Fig.~\ref{fig:rmh_map} is to validate that the TEOM framework correctly handles a $\rho$-dependent generator $\mathcal{L} = \mathcal{L}[f,\rho]$ and reproduces the interaction-induced qualitative evolution of the spectrum, including the chain-rule contribution in Eq.~\eqref{eq:DL_chainrule}.
We find that the MFD results capture the qualitative features obtained by iTEBD.
Specifically, as $U$ increases, the spectral weight of the positive continuum along $\omega_2 = \omega_1$ accumulates toward $\omega_1 \approx 2$, and the sign of $\bar{\chi}^{(2)}(\omega_1, \omega_2)$ changes along $\omega_2 = -\omega_1$ in the range $1 \lesssim \vert \omega_1 \vert \lesssim 2$.
Importantly, if MFD is not taken into account, namely, if the mean-field values are kept fixed at their initial-state values, the interaction effect only slightly renormalizes $h_z$, and the frequency-dependent changes in $\bar{\chi}^{(2)}(\omega_1, \omega_2)$ are not captured.
In this sense, these features can be attributed to many-body effects beyond the one-body physics.

\begin{figure}[t]\centering
\includegraphics[scale=1]{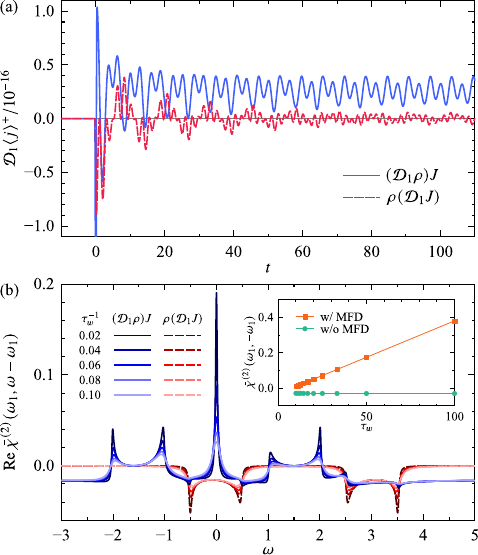}
\caption{Decomposition of the first Gateaux derivative into two contributions [Eq.~\eqref{eq:current_decomposition}].
(a) Temporal profiles of the electric current, showing the contributions from the change in the state, $\propto \Tr[(\mathcal{D}_1 \rho) J]$ (blue solid), and the change in the band structure, $\propto \Tr[\rho (\mathcal{D}_1 J)]$ (red dashed).
(b) Second-order response function plotted as a function of the response frequency $\omega = \omega_1 + \omega_2$, for different values of the temporal width of the exponential window function, $\tau_w$.
The blue and red lines show the $(\mathcal{D}_1 \rho) J$ and $\rho (\mathcal{D}_1 J)$ contributions, respectively.
The inset in panel (b) shows the dependence of $\bar{\chi}^{(2)}(\omega_1,-\omega_1)$ on $\tau_w$; results with MFD are indicated by squares, whereas those without MFD are indicated by circles.
In both panels (a) and (b), the interaction strength is fixed at $U = 0.1$, and the perturbation field $g_{\alpha_1\!} = g_{\text{cos}_1\!}$ with $\omega_1 = 1.5$ is used.
}
\label{fig:rmh_decomp}
\end{figure}

We now discuss in more detail the optical-rectification response $\bar{\chi}^{(2)}(\omega_1, -\omega_1)$ for $1 \lesssim \vert \omega_1 \vert \lesssim 2$.
Motivated by the qualitative agreement between MFD and iTEBD in the frequency-resolved spectrum (Fig.~\ref{fig:rmh_map}), we use the TEOM-based term decomposition below to identify which part of the MFD response is robust and which part is susceptible to approximation-specific artifacts.
Figure~\ref{fig:rmh_decomp}(a) shows the time dependence of the two contributions to the first Gateaux derivative of the current [Eq.~\eqref{eq:current_decomposition}], for $\omega_1 = 1.5$.
The contribution $(\mathcal{D}_1 \rho) J$, namely, the current arising from the state variation induced by the perturbation-field excitation, is triggered at $t = 0$ and then continues to flow with a nonzero mean value.
In contrast, the contribution $\rho (\mathcal{D}_1 J)$, namely, the current arising from the change in the current operator induced by the perturbation field, becomes negative shortly after $t = 0$ and subsequently exhibits oscillations around a zero mean from $t \sim 10$ onward.
This behavior is consistent with the standard time-domain signatures of an injection-current-like (long-lived dc) component for the former and a shift-current-like (zero-mean oscillatory) component for the latter \cite{Belinicher1980, Sipe2000, VonBaltz1981, Young2012}.
In what follows, we quantify this distinction via the $\tau_w$ dependence and use it as a term-resolved diagnostic.

Figure~\ref{fig:rmh_decomp}(b) shows $\re \bar{\chi}^{(2)}(\omega_1, \omega-\omega_1)$ decomposed into the two contributions in Eq.~\eqref{eq:current_decomposition}, after applying the exponential window of width $\tau_{w}$.
Here, $\tau_{w}$ plays the role of an effective long-time cutoff and can be viewed as a phenomenological relaxation time in the frequency extraction.
Consistent with its nonzero long-time average in Fig.~\ref{fig:rmh_decomp}(a), the contribution from $(\mathcal{D}_1 \rho) J$ (blue solid lines) yields a low-frequency enhancement whose magnitude grows with $\tau_{w}$ at $\omega = 0$.
In contrast, the contribution from $\rho (\mathcal{D}_1 J)$ (red dashed lines) is essentially insensitive to $\tau_w$ near $\omega = 0$.

The inset plots $\bar{\chi}^{(2)}(\omega_1,-\omega_1)$, which is real by Eqs.~\eqref{eq:chi_conjugate} and \eqref{eq:symmetrized_chi}, as a function of $\tau_w$.
Without MFD, $\bar{\chi}^{(2)}(\omega_1,-\omega_1)$ remains constant, whereas with MFD it increases approximately linearly with $\tau_w$.
A similar $\tau_{w}$-dependent feature was also observed in mean-field-based nonlinear optical spectra and was attributed to collective excitations \cite{Kaneko2021}.
In the present nonlinear setting, we find that the entire $\tau_w$-sensitive dc-like component can be traced to the state-variation channel $(\mathcal{D}_1 \rho)J$, while the explicit operator-variation channel $\rho(\mathcal{D}_1 J)$ remains well behaved.
Taken together with Fig.~\ref{fig:rmh_map}, this indicates that MFD can capture the dominant interaction-induced spectral rearrangements in frequency-resolved response functions, in qualitative agreement with iTEBD, while the TEOM decomposition pinpoints the specific channel responsible for the potentially spurious long-lived dc component.
This term-resolved separation is useful as a diagnostic:
Injection-current-like residual dc responses are known to be artificially generated by time-dependent mean-field approximations under linearly polarized driving in time-reversal-symmetric settings, and can dominate the intrinsic shift-current contribution through a stronger low-frequency divergence \cite{Sato2024}.
Our TEOM-based decomposition makes explicit which contribution carries such sensitivity, thereby enabling controlled interpretation and targeted improvement of approximate dynamics.

\subsection{Higher-order response functions} \label{sec:higher-order}
We consider a two-dimensional four-band model, which serves as an example of a more realistic solid-state electron system, and we compute the response functions up to fifth order and discuss high-harmonic generation in the perturbative regime.

As a concrete realization, we study a ferromagnetic Kondo lattice model with a four-sublattice all-in--all-out spin configuration.
This spin structure is characterized by a spatially uniform scalar chirality, which endows the electronic bands with nonzero Chern numbers \cite{Martin2008, Akagi2010}.
Therefore, this state can be regarded as a skyrmion crystal state with the smallest magnetic unit cell and has recently been reported in $\mathrm{CoTa_{3}S_{6}}$ and $\mathrm{CoNb_{3}S_{6}}$ \cite{Takagi2023, Park2023}.
The Hamiltonian matrix can be written as
\begin{align}
H(\bm{k}) &= -\frac{J_{\mathrm{K}}}{\sqrt{3}} \sigma_z \otimes \sigma_0
+ \frac{J_{\mathrm{K}}}{\sqrt{3}} (\sigma_x+\sigma_y) \otimes \sigma_y \notag \\
&\quad - 2h_1\cos(k_x a)\, \sigma_0 \otimes \sigma_x \notag \\
&\quad - 2h_1\cos\left(\frac{k_x+\sqrt{3}k_y}{2}a\right)\, \sigma_x \otimes \sigma_0 \notag \\
&\quad - 2h_1\cos\left(\frac{k_x-\sqrt{3}k_y}{2}a\right)\, \sigma_x \otimes \sigma_x,
\end{align}
where $h_1$ is the nearest-neighbor transfer integral, $J_{\mathrm{K}}$ is the exchange interaction between the electrons and localized spins, and $\sigma_{\alpha}$ denotes the Pauli matrix (with the identity matrix for $\alpha = 0$).
The electric current density is defined by $\langle \bm{j} \rangle = 2(NV)^{-1} \sum_{\bm{k}} \Tr[ \rho \partial H(\bm{k} - e\bm{A}/\hbar)/\partial \bm{k}]$, where the prefactor $2$ accounts for a symmetry-related twofold degeneracy, $V = 2\sqrt{3} a^2$ is the area of the magnetic unit cell ($a$ is the nearest-neighbor distance), and $N$ is the number of $\bm{k}$ points.
The vector potential $\bm{A}(t) = \bm{f}(t)$ is given by
\begin{align}
\bm{A}(t) = \bm{F}_0 u_{\tau_{f}\!}(t), \quad
\bm{F}_0 = (F_0 \cos\psi, F_0 \sin\psi),
\end{align}
where $\psi$ is a polarization angle measured from the $x$ axis.
See Ref.\ \cite{Ono2024} for details of the model.
Hereafter, we set $h_1 = a = e = \hbar = 1$ and $N = 256^2$.
We take the initial state to be the insulating ground state at half filling for $J_{\mathrm{K}} = 3$.

For a quantitative comparison with direct cw simulations, in this section we introduce dissipation via the relaxation-time approximation (RTA).
Specifically, Eq.~\eqref{eq:vonNeumann} is modified as follows:
\begin{align}
\frac{\mathrm{d}}{\mathrm{d}t} \rho(\bm{k}) = -\frac{\mathrm{i}}{\hbar} [H(\bm{k} - \bm{A}), \rho(\bm{k})] - \gamma [\rho(\bm{k}) - \rho_{0}(\bm{k})].
\label{eq:vonNeumann_RTA}
\end{align}
The second term represents phenomenological dissipation, where $\gamma$ is the dissipation rate and $\rho_0(\bm{k})$ denotes the density matrix in the ground state of $H(\bm{k})$.
The Gateaux derivative of Eq.~\eqref{eq:vonNeumann_RTA} is given by
\begin{align}
\frac{\mathrm{d}}{\mathrm{d}t} \mathcal{D}_{J} \rho(\bm{k}) = -\frac{\mathrm{i}}{\hbar} \mathcal{D}_{J} [H(\bm{k} - \bm{A}), \rho(\bm{k})] - \gamma \mathcal{D}_{J} \rho(\bm{k})
\label{eq:vonNeumann_RTA_TEOM}
\end{align}
for $J \neq \varnothing$, where we assume that $\rho_{0}(\bm{k})$ is independent of the vector potential.
In general, $\rho_0$ is taken to be the ground-state density matrix evaluated at the shifted wave vector $\bm{k} - \bm{A}(t)$.
Here, however, we instead use the ground-state density matrix at $\bm{A} = 0$ for each $\bm{k}$, i.e., $\rho_0(\bm{k})$.
This approximation is reasonable because the delta-function-like external field $\bm{A} = \bm{f}$ is nonzero only within a short time window $\vert t \vert \lesssim \tau_f \ll \gamma^{-1}$.
With this choice, we avoid diagonalizing $H(\bm{k}-\bm{A})$ to obtain $\rho_0$ at each time step, and we also eliminate the need to consider the Gateaux derivatives of $\rho_0$, which are not straightforward to evaluate because $\rho_0(\bm{k} - \bm{A})$ is not evolved by the EOM.

To calculate the nonlinear response functions, we set the temporal width of the vector potential to $\tau_f = 0.01$ to cover the required energy range of the system.
We use a time step of $\delta t = 2^{-9}$ for $\lvert t \rvert < 6\tau_f$ and $\delta t = 2^{-6}$ otherwise.
For the perturbation field $g(t)$ and the window function $w(t)$, we take $\tau_g \to \infty$ and $\tau_w \to \infty$ when dissipation is included ($\gamma > 0$), whereas for the dissipationless case ($\gamma = 0$) we use a Hann envelope\footnote{Because the Hann window has finite support in the time domain, it reduces the computational cost.
In the present calculations, for the Hann window we simulate the time range $\lvert t \rvert \leq 2\tau_w = 20$, whereas for the exponential window we require approximately $\lvert t \rvert \leq 80$.} with widths $\tau_g = \tau_w = 10$.

We begin by verifying that the nonlinear response functions computed within our framework yield the correct results by comparing them with those obtained using an alternative approach.
In this approach, which we refer to here as the cw-driving method, the nonlinear response functions are obtained by extracting the amplitude and phase of each harmonic from the Fourier spectra of the currents in the steady state under cw driving (see Appendix~\ref{sec:cw} for details).

\begin{figure*}[t]\centering
\includegraphics[scale=1]{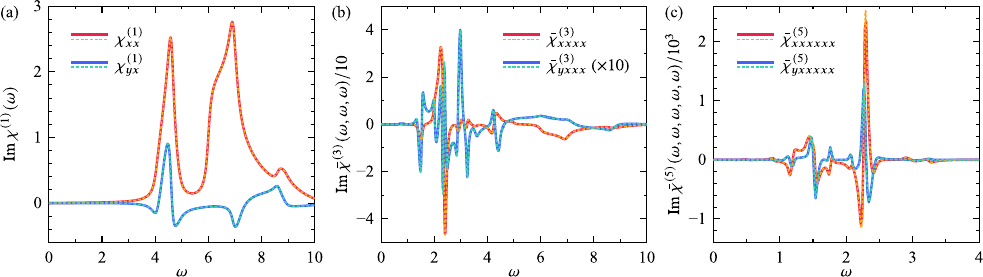}
\caption{Linear and nonlinear response functions for harmonic generation in the two-dimensional four-band model.
(a) First-order, (b) third-order, and (c) fifth-order response functions.
The results obtained using TEOM are shown as solid lines, and those obtained using the Kubo formula (for $n = 1$) and continuous-wave driving (for $n = 3, 5$) are shown as dashed lines.
}
\label{fig:trig_chi135}
\end{figure*}

Figure~\ref{fig:trig_chi135} shows the first-, third-, and fifth-order harmonic response functions, $\chi^{(1)}(\omega)$, $\bar{\chi}^{(3)}(\omega,\omega,\omega)$, and $\bar{\chi}^{(5)}(\omega,\omega,\omega,\omega,\omega)$, for $\gamma = 0.1$.
The dashed lines indicate the corresponding response functions obtained from the Kubo formula for $n = 1$, and from the cw-driving method for $n = 3$ and $5$.
Here, we set the polarization angle $\psi = 0$ and compute the response functions from the induced current densities in the $x$ and $y$ directions, which can be simultaneously calculated.
For all three orders shown, the results obtained from the TEOM agree well with those from the Kubo formula and the cw-driving method, accurately reproducing the excitation structure.
Because the cw-driving protocol probes steady-state harmonic generation, whereas TEOM reconstruct response kernels from infinitesimal functional derivatives, their agreement provides a methodologically independent validation of the TEOM calculations up to the fifth order.
In particular, diagonal contractions of the fifth-order kernel (e.g., the cut $\omega_5 = \omega_1$ used below) reduce to the fifth-harmonic response and are cross-checked against the cw-driving results.

\begin{figure*}[t]\centering
\includegraphics[scale=1]{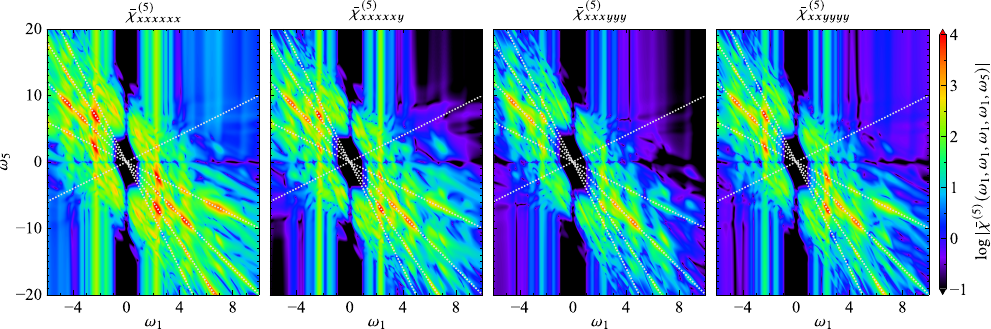}
\caption{Color maps of the absolute values of the fifth-order response functions, $\bar{\chi}_{xxxxxx}^{(5)}$, $\bar{\chi}_{xxxxxy}^{(5)}$, $\bar{\chi}_{xxxyyy}^{(5)}$, and $\bar{\chi}_{xxyyyy}^{(5)}$ (left to right), in the two-dimensional four-band model with $\omega_1 = \omega_2 = \omega_3 = \omega_4$.
The dotted lines indicate $\omega_5 = \omega_1$, $-\omega_1$, $-2\omega_1$, $-3\omega_1$, and $-4\omega_1$.
}
\label{fig:trig_chi5}
\end{figure*}

Having validated the diagonal harmonic responses, we now consider the additional information obtained from the same single-frequency scan.
The TEOM results shown in Fig.~\ref{fig:trig_chi135} are obtained from a scan over a single perturbation-field frequency\footnote{This scan also provides components with different sign patterns, e.g., $\bar{\chi}^{(5)}(\omega_1,\omega_1,\omega_1,-\omega_1, \omega - 2\omega_1)$ and $\bar{\chi}^{(5)}(\omega_1,\omega_1,-\omega_1,-\omega_1, \omega)$.}, $\omega_1 = \omega_2 = \omega_3 = \omega_4$.
The four independent components obtained from the scan are shown in Fig.~\ref{fig:trig_chi5}.
To the best of our knowledge, Fig.~\ref{fig:trig_chi5} provides the first two-dimensional frequency-resolved slices of a fifth-order response function for a solid-state electron system, namely, $\bar{\chi}^{(5)}(\omega_1,\omega_1,\omega_1,\omega_1,\omega_5)$ as a function of $(\omega_1,\omega_5)$.
While standard cw driving directly yields only diagonal slices (harmonic responses), the present TEOM-based protocol reconstructs this kernel slice and returns the full $\omega_5$ dependence for each scanned $\omega_1$.
The same approach can target other constrained slices by choosing the perturbation-field frequencies accordingly, at a corresponding computational cost.
Moreover, evaluating fifth-order response functions with earlier approaches, either using existing methods (e.g., diagrammatic techniques) or using a finite-difference approximation \cite{Ono2025}, was impractical because of prohibitive computational costs or severe subtractive-cancellation errors, respectively.

In Fig.~\ref{fig:trig_chi5}, the intensity vanishes near $\omega_1 = \omega_5 = 0$ because the system considered is an insulator.
The white dotted line $\omega_5 = \omega_1$ indicates the cut used to obtain the fifth-harmonic response shown in Fig.~\ref{fig:trig_chi135}(c).
We also find that strong peaks (continua) appear along the lines $\omega_5 = -m\omega_1$ ($m=1,2,3,4$).
The strongest intensity appears around $\omega_1 \sim 2.3$ for $\omega_5 = -3\omega_1$ (i.e., $\omega = \omega_1$) and $\omega_5 = -\omega_1$ (i.e., $\omega = 3\omega_1$).
Overall, the intensity is strongest for $\bar{\chi}^{(5)}_{xxxxxx}$, while the other off-diagonal components are weaker.

In addition, a broad peak appearing at $\omega_1 \sim 6\text{--}7$ and $\omega_5 \sim 2\text{--}3$ lies off the harmonic-response line ($\omega_5 = \omega_1$).
Although it is not prominent in $\bar{\chi}^{(5)}_{xxxxxx}$ and $\bar{\chi}^{(5)}_{xxxyyy}$, it emerges as a relatively pronounced feature in $\bar{\chi}^{(5)}_{xxxxxy}$.
Given this frequency range, we infer that the main contributions to this structure are a one-photon excitation to the upper band at $\omega \sim 6\text{--}7$ in $\chi^{(1)}$ in Fig.~\ref{fig:trig_chi135}(a), and a one-photon excitation between that band and the lower band at $\omega \sim 4\text{--}5$.
In this way, we expect that frequency-resolved higher-order response functions enabled by our framework will provide access to richer dynamical and spectroscopic properties of the physical system.

\begin{figure}[t]\centering
\includegraphics[scale=1]{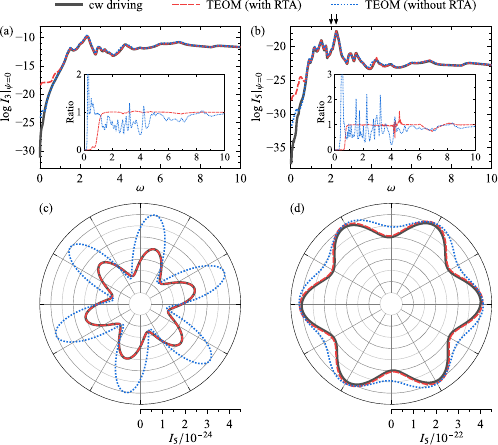}
\caption{Harmonic intensities computed using the cw-driving method (gray solid lines) and using the TEOM, either with no window function for $\gamma = 0.1$ (red dashed lines) or with a Hann window for $\gamma = 0$ (blue dotted lines).
(a), (b) Intensities of (a) the third and (b) the fifth harmonics, with the polarization angle $\psi = 0$.
The insets show the ratio of the cw-driving result to each TEOM result.
Polarization-angle dependence of the fifth-harmonic intensity for (c) $\omega = 2.011$ and (d) $2.262$; these frequencies are indicated by the arrows in panel (b).
}
\label{fig:trig_hhg}
\end{figure}

To further validate the fifth-order response functions shown above, and to present an example in which distinctive features first emerge only at fifth order or higher, we discuss the polarization-angle $\psi$ dependence of the fifth-harmonic intensity in the perturbative regime.
Because this observable depends on a coherent (phase-sensitive) superposition of multiple off-diagonal components of the fifth-order response tensor, its $\psi$ dependence provides a particularly stringent check of the calculation.
When electrons are driven by a cw field with frequency $\omega_{\mathrm{cw}}$ and reach a steady state, the Fourier transform of the current density yields the amplitude $\bm{A}_n$ and phase $\phi_n$ of the $n$th harmonic in $\langle j_\alpha \rangle(t) = \sum_{n\geq 0} A_{n\alpha} \cos(n\omega_{\mathrm{cw}}t - \phi_{n\alpha})$.
The radiated intensity of the $n$th harmonic is proportional to
\begin{align}
I_n = \frac{(n\omega_{\mathrm{cw}})^2}{2} \Vert \bm{A}_{n} \Vert^2. \label{eq:In_cw}
\end{align}
Alternatively, given the $n$th-order response functions, $I_n$ can be calculated through Eqs.~\eqref{eq:I1}--\eqref{eq:I5} in Appendix~\ref{sec:In}.
For the present system with sixfold rotational symmetry, these equations imply that $I_1$ and $I_3$ are isotropic and therefore independent of the polarization angle $\psi$, whereas $I_5$ exhibits a characteristic angular dependence through $\cos(6\psi)$ and $\sin(6\psi)$.
Confirming agreement between these two evaluations would further support the validity of our framework.

In Figs.~\ref{fig:trig_hhg}(a) and \ref{fig:trig_hhg}(b), we show the third- and fifth-harmonic intensities for $\psi = 0$.
The thick gray lines show the results from the cw-driving method [Eq.~\eqref{eq:In_cw}].
The red dashed lines correspond to TEOM with the RTA ($\gamma = 0.1$) and no window function, while the blue dotted lines correspond to TEOM without the RTA ($\gamma = 0$) but with Hann-type envelope and window functions ($\tau_g = \tau_w = 10$).
These TEOM results agree well with the cw-driving results, reproducing the peak structure.

The insets show the ratio of the cw-driving result to the corresponding TEOM result, $r_{n}(\omega) = (I_n^{\text{cw}} / I_n^{\text{TEOM}}) \vert_{\psi=0}$.
For TEOM with the RTA, the ratio is close to unity and the agreement is excellent, except in the subgap region $\omega \lesssim 1$.
In contrast, the TEOM results without the RTA reproduce the overall spectral structure down to relatively low energies, $\omega \gtrsim 0.5$, but their absolute values show an $\mathcal{O}(1)$ discrepancy from the cw-driving results.

We show the polarization-angle dependence of $I_5$ for $\omega = 2.011$ and $2.262$ in Figs.~\ref{fig:trig_hhg}(c) and \ref{fig:trig_hhg}(d).
For comparison, the Hann-window TEOM data $I_5^{\text{TEOM}}$ are multiplied by $r_5(2.011) = 0.679$ in panel (c) and $r_5(2.262) = 1.76$ in panel (d), so that they match the cw-driving values at $\psi = 0$.
In both cases, the TEOM results with the RTA almost perfectly reproduce the cw-driving result.
Importantly, the fifth-harmonic polarization anisotropy, which is highly sensitive to phase and interference among multiple pathways, is captured quantitatively (within numerical accuracy) without introducing any adjustable parameters.
In contrast, the Hann-window results (without the RTA) either overestimate or underestimate the $\psi$-dependent oscillatory component even after this rescaling and show a slight shift in the lobe direction.

These results demonstrate that the harmonic-generation intensity in the perturbative regime can be explained quantitatively by the nonlinear response function obtained using the TEOM.
They also indicate that, for direct comparisons with RTA-based simulations, TEOM calculations should likewise incorporate the RTA, and the temporal width of the perturbation fields should be taken to the cw limit ($\tau_g \to \infty$).

Finally, we comment on the computational cost.
The calculations of the fifth-order response functions were performed on the ISSP supercomputer system at the University of Tokyo.
Each node is equipped with two AMD EPYC 7702 (2.0 GHz) CPUs with 64 physical cores each.
For a given $\omega_1$, there are $2^4 = 16$ independent cosine/sine perturbation-field combinations; each combination corresponds to a single task (i.e., one of the 16 runs).
We used an embarrassingly parallel task-level scheme across nodes, without the Message Passing Interface (MPI), and executed each task on a single node with eight threads.
With 128 physical cores per node, this configuration allows up to 16 tasks to run concurrently per node, and up to 384 tasks concurrently on 24 nodes, subject to scheduler allocation.
Under the calculation conditions used in the main text, the average wall-clock time was 42.7 min per task.
The full $\omega_1$ scan used to generate one panel of Fig.~\ref{fig:trig_chi5} comprised 576 points in $\omega_1$ (from $0$ to $11.5$ in steps of $0.02$), corresponding to 9216 independent tasks, each involving time evolution over $\vert t \vert \leq 20$.
By running these tasks concurrently on 24 nodes, the full scan took approximately 17.5 h in total, bringing the fifth-order computation into a practically feasible regime with a modest high-performance computing allocation.
Note that the total wall-clock time is expected to scale approximately inversely with the number of concurrently executed tasks and thus roughly inversely with the number of available nodes.

\subsection{Classical Duffing oscillator} \label{sec:duffing}
The numerical examples presented in the previous sections have focused on quantum solid-state electron systems.
In this section, we consider a completely different physical system, the Duffing oscillator (a classical oscillator with a quartic anharmonic potential), which serves as a representative model of weak anharmonicity and nonlinear response in diverse contexts, including mechanics \cite{Aldridge2005, Rhoads2010}, electronics \cite{Vijay2009}, and acoustics \cite{Yu2011, Fang2025}.
In such resonant systems, weak quartic anharmonicity is often well captured at leading order by a Duffing-type cubic restoring-force term \cite{Nayfeh2008, Krantz2019, Ueda1991}.
We use this system to demonstrate the generality of our framework and to illustrate how the computational time and error scale with the response order $n$.

The EOM of the Duffing oscillator is given by
\begin{align}
\ddot{x}(t) + \gamma \dot{x}(t) + \omega_0^2 x(t) + \alpha x(t)^3 = f(t),
\label{eq:duffing}
\end{align}
where $\omega_0$ is the natural angular frequency of the corresponding linear (harmonic) oscillator, $\gamma$ ($> 0$) is the linear damping coefficient, $\alpha$ is the nonlinearity coefficient, and $f(t)$ is the external driving force.
The displacement $x$ is a functional of the external driving force, i.e., $x = x[f]$.
We recast Eq.~\eqref{eq:duffing} into an equivalent system of ordinary differential equations by introducing the velocity $v(t) = \dot{x}(t)$, which yields
\begin{align}
\frac{\mathrm{d}}{\mathrm{d}t} x = v, \quad
\frac{\mathrm{d}}{\mathrm{d}t} v = f - \gamma v - \omega_0^2 x - \alpha x^3.
\label{eq:duffing_xv}
\end{align}

The Gateaux derivatives $\mathcal{D}_{J}$ of these coupled equations are given by
\begin{gather}
\frac{\mathrm{d}}{\mathrm{d}t} \mathcal{D}_{J} x = \mathcal{D}_{J} v, \\
\frac{\mathrm{d}}{\mathrm{d}t} \mathcal{D}_{J} v = \mathcal{D}_{J}f - \gamma \mathcal{D}_{J}v - \omega_0^2 \mathcal{D}_{J}x - \alpha \mathcal{D}_{J}x^3,
\label{eq:duffing_teom}
\end{gather}
where $J$ is an index set labeling the perturbation fields [see Eq.~\eqref{eq:def_DI}].
In our protocol for extracting the $n$th-order response, we take $n-1$ Gateaux derivatives, so that $J \subseteq I_{n-1}$.
Accordingly, there are $2^{n-1}$ derivative components for each of $x$ and $v$, i.e., $2\times 2^{n-1} = 2^{n}$ dynamical variables in total.
Since the driving force $f$ enters linearly, only the first Gateaux derivative $\mathcal{D}_{g_{\alpha_i\!}} f(t) = g_{\alpha_i\!}(t)$ is nonzero, whereas its higher-order derivatives vanish.

The Gateaux derivative of the anharmonic term $x^3$ can be evaluated as a sum over ordered partitions of the index set $J$ into three (possibly empty) disjoint subsets,
\begin{align}
\mathcal{D}_{J} x^3 = \sum_{A \sqcup B \sqcup C = J} (\mathcal{D}_{A} x) (\mathcal{D}_{B} x) (\mathcal{D}_{C} x).
\label{eq:DJ_x3}
\end{align}
In the fully general TEOM formulation, direct enumeration of this partition sum scales exponentially in $\vert J \vert$.
However, in the symmetric setting where all perturbation fields are identical and $\mathcal{D}_J$ depends only on the cardinality $k = \vert J \vert$, the set-indexed derivatives collapse to a single-index sequence $\{ \mathcal{D}^k x \}_{k=0}^{n-1}$.
The same partitions can then be counted combinatorially, yielding the multinomial form
\begin{align}
\mathcal{D}^k x^3 = \sum_{a+b+c = k} \frac{k!}{a! b! c!} (\mathcal{D}^a x) (\mathcal{D}^b x) (\mathcal{D}^c x),
\label{eq:DJ_x3_sym}
\end{align}
so the cubic term can be evaluated in $\mathcal{O}(n^2)$ time via an exponential generating function (equivalently, two Cauchy products).

We calculate the $n$th-order response function
\begin{align}
\bar{\chi}^{(n)}(\omega_1,\dots,\omega_n)
= \frac{(2\pi)^{n-1}}{f(\omega_n)} \frac{\delta^{n-1} x^{(n)}(\omega)}{\delta f(\omega_{n-1}) \cdots \delta f(\omega_{1})},
\label{eq:chi_duffing}
\end{align}
using the complex-valued perturbation fields in Eq.~\eqref{eq:gexp_t}.
We set the perturbation-field frequencies to $\omega_1 = \omega_2 = \cdots = \omega_{n-1}$ and scan over the interval $0 \leq \omega_1 \leq 2$ with a step of $0.002$.
Since inversion symmetry eliminates even-order responses in this system, we can replace the $n$th-order displacement $x^{(n)}(\omega)$ in Eq.~\eqref{eq:chi_duffing} with the total response $x(\omega)$.
We set $\omega_0 = 1$ as the unit of frequency, and use the parameters $\gamma = 0.1$, $\alpha = 0.1$, $F_0 = 10^{-4}$, $\tau_f = 0.01$, $\tau_g \to \infty$ (continuous-wave limit), and $\tau_w \to \infty$ (no window function).
We perform the numerical integration using the fourth-order Runge--Kutta method with a time step of $\delta t = 2^{-14}$ for $|t| < 6\tau_f$ and $\delta t = 2^{-10}$ otherwise, and we simulate the dynamics from $t = -4000$ to $4000$.

\begin{figure}[t]\centering
\includegraphics[scale=1]{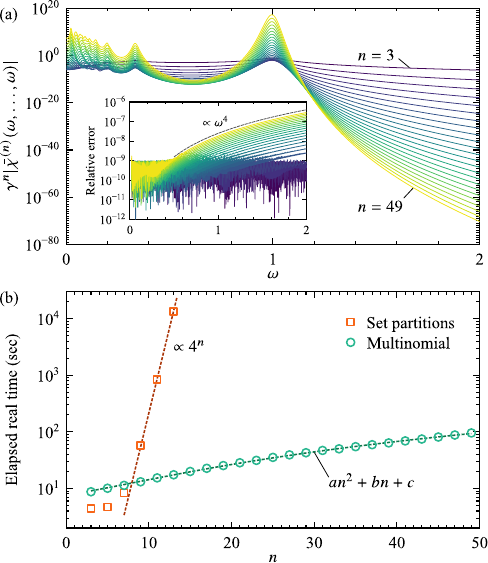}
\caption{Nonlinear response functions of the Duffing oscillator.
(a) Magnitudes of the $n$th-order response functions $\bar{\chi}^{(n)}(\omega,\dots,\omega)$ multiplied by $\gamma^n$ for $n = 3, 5, \dots, 49$ (dark-to-light colors).
The inset shows the relative error with respect to the Green's-function results.
(b) Elapsed real time to compute $\bar{\chi}^{(n)}(\omega_1,\dots,\omega_1,\omega-(n-1)\omega_1)$ for a single input frequency $\omega_1$, as a function of the response order $n$, using explicit ordered set-partition enumeration in Eq.~\eqref{eq:DJ_x3} (squares) and the size-reduced multinomial evaluation in Eq.~\eqref{eq:DJ_x3_sym} (circles).
The fitted functions $\propto 4^n$ (orange) and $an^2 + bn + c$ with $a = 0.028$, $b = 0.38$, and $c = 7.67$ (green) are also shown.
}
\label{fig:duffing}
\end{figure}

Figure \ref{fig:duffing}(a) shows the computed $n$th-order response function $\bar{\chi}^{(n)}(\omega,\dots,\omega)$ corresponding to the harmonic responses.
These results are obtained by setting the input frequencies to $\omega_1 = \dots = \omega_{n-1}$ and varying $\omega_1$ from $0$ to $2$ in steps of $0.002$.
Since the system is a classical model with a single degree of freedom, we can systematically and accurately compute responses up to $n = 49$.
The largest peak occurs at $\omega = 1$, and several additional peaks appear at $\omega = 1/k$ with $k = 3,5,\dots,n$ for the $n$th-order response, corresponding to multi ``photon'' absorption processes.
The inset of Fig.~\ref{fig:duffing}(a) shows the relative error with respect to the exact results obtained by the Green's-function method (see Appendix~\ref{sec:duffing_green}).
Importantly, the relative error is below $10^{-9}$ for $n \leq 13$, and it remains below $10^{-6}$ even for higher orders up to $n = 49$ over this frequency range.
This level of accuracy is possible because the Gateaux derivative can rigorously isolate the contribution at each order.
In addition, for higher orders with $n \geq 15$, the relative error shows an increase proportional to $(\omega \delta t)^4$, which is attributable to the present Runge--Kutta method error scaling with the fourth power of the time step $\delta t$.

Figure~\ref{fig:duffing}(b) plots the elapsed wall-clock time\footnote{For this benchmark, we performed separate simulations for each order $n$ to quantify the run-time as a function of $n$.
In practical applications, integrating the TEOM hierarchy up to a maximum order $N$ propagates all derivatives up to $k = N-1$ simultaneously
(e.g., $\{\mathcal{D}^k x\}_{k=0}^{48}$ for $N = 49$), so the response functions for all lower orders $n < N$ can be obtained from a single run.} required to compute $\bar{\chi}^{(n)}(\omega_1,\dots,\omega_1,\omega-(n-1)\omega_1)$ for a single input frequency $\omega_1$ as a function of the response order $n$.
The calculations were performed on a single workstation equipped with one AMD EPYC 9654 (2.4~GHz) CPU, where we evaluated 96 values of $\omega_1$ concurrently using one core per task.
Using the fully general TEOM with the ordered set partitions in Eq.~\eqref{eq:DJ_x3} (squares), the run-time is dominated by a fixed computational overhead for moderate orders ($n \approx 3\text{--}7$), and grows exponentially for higher orders, consistent with $\mathcal{O}(4^n)$ for $n=9\text{--}13$.
This behavior reflects the combinatorial cost of evaluating $\mathcal{D}_J x^3$ by direct enumeration over ordered partitions together with the proliferation of set-indexed derivatives in the general TEOM (see Sec.~\ref{sec:discussion} for details).
In contrast, in the symmetric setting where the set-indexed derivatives collapse to the cardinality-indexed sequence $\{\mathcal{D}^k x\}_{k=0}^{n-1}$ and the cubic term is evaluated by the multinomial form in Eq.~\eqref{eq:DJ_x3_sym} (circles), the run-time follows the quadratic fit $an^2+bn+c$, i.e., $\mathcal{O}(n^2)$.
Note that the crossover between the quadratic and linear contributions occurs at $n \sim b/a \approx 13.7$, so for the present range ($n \lesssim 49$) the linear term $bn$ is not negligible even though the asymptotic scaling is $\mathcal{O}(n^2)$.
Taken together, these results show that while the fully general implementation has an exponential scaling in $n$, the symmetric formulation enables efficient high-order calculations in the diagonal-frequency configuration, and in both cases the TEOM framework avoids the severe cancellation errors of finite differencing and the explicit diagrammatic enumeration of multipoint correlators.

\section{Discussion} \label{sec:discussion}
To clarify the computational cost of our approach, it is useful to distinguish the order $n$ of the response function from the order $m$ of the Gateaux derivative used in the protocol.
In the weak-source frequency-domain extraction based on Eq.~\eqref{eq:chi_nth}, the $(n-1)$th Gateaux derivative eliminates all lower-order response contributions, so that the target $n$th-order response is isolated by taking $m=n-1$ derivatives [after (anti)symmetrization in Eq.~\eqref{eq:symmetrized_Q}].
In the zero-field source-derivative formulation introduced in Sec.~\ref{sec:zero_field}, the broadband source pulse is instead included as an additional Gateaux-derivative direction and the reconstruction is performed at $f=0$; hence, the corresponding derivative order is $m=n$.
The numerical demonstrations in Sec.~\ref{sec:demo} use the weak-source formulation.
The TEOM hierarchy for derivatives up to order $m$ consists of the set $\{\mathcal{D}_{J} \rho\}_{J \subseteq I_m}$ [Eq.~\eqref{eq:Dn_rho}], whose cardinality is $2^m$.
Thus, the number of dynamical variables to be propagated grows as $\mathcal{O}(2^m)$, with $m=n-1$ for the weak-source formulation and $m=n$ for the zero-field source-derivative formulation.

However, the wall-clock time is not determined by the factor $2^m$ alone; it is dominated by the cost of evaluating the right-hand side of the TEOM hierarchy.
The key structure is the subset expansion in Eq.~\eqref{eq:Dn_rho}, where $\mathrm{d} \mathcal{D}_{J}\rho/\mathrm{d}t$ couples to $\mathcal{D}_{J\setminus K} \rho$ through $\mathcal{D}_{K} \mathcal{L}$ summed over all subsets $K \subseteq J$.
If the Liouville operator $\mathcal{L}$ depends linearly on the external field (e.g., via a linear coupling such as a Zeeman term in spin systems), then $\mathcal{D}_{K} \mathcal{L} = 0$ for $\vert K \vert \geq 2$, and the subset sum collapses to contributions from $K = \varnothing$ and singleton sets.
In that common situation, the per-step algebraic work is only polynomially larger than $2^m$.
In contrast, when the field enters $\mathcal{L}$ nonlinearly, as discussed in Secs.~\ref{sec:RM}--\ref{sec:higher-order}, higher-order derivatives $\mathcal{D}_{K} \mathcal{L}$ do not vanish, and the full subset sum must be retained at every order.
In such cases, a naive evaluation of the subset couplings leads to an additional combinatorial factor, and the total per-step complexity of the hierarchy can scale as $\mathcal{O}(3^m)$ [see Eq.~\eqref{eq:subset_couplings}], up to model-dependent prefactors reflecting the cost of constructing and applying $\mathcal{D}_{K} \mathcal{L}$.

A further increase can occur when the EOM is nonlinear in the dynamical variables (i.e., $\mathcal{L} = \mathcal{L}[f, \rho]$ with polynomial-type nonlinearities), because $\mathcal{D}_{J} \mathcal{L}$ then contains partition sums generated by repeated application of the Leibniz (product) rule.
As an illustrative example, in our Duffing-oscillator implementation the nonlinear force contains a cubic term $x^3$, and its derivative is evaluated according to Eq.~\eqref{eq:DJ_x3}, which involves a sum over ordered partitions of the index set $J$.
Since the number of ordered partitions of $J$ into three subsets is $3^{\vert J \vert}$, summing this cost over all $2^m$ subsets yields $\sum_{k=0}^{m} \binom{m}{k} 3^k = 4^m$.
This explains why the Duffing case exhibits an apparent $\mathcal{O}(4^m)$ scaling even though the underlying TEOM hierarchy contains only $2^m$ dynamical components.

These exponential scalings should be contrasted with earlier approaches such as diagrammatic expansions, which typically become prohibitive at high orders not only because of factorial or super-exponential algebraic complexity, but also because each order requires integral evaluations that must be constructed separately for each topology.
In contrast, the TEOM hierarchy is generated systematically and uniformly once the EOM is specified, regardless of the target response order.

With this comparison in mind, we next discuss the relationship between the TEOM framework and alternative derivative-evaluation engines such as AD \cite{Griewank2008}.
Because the reconstruction protocol in Sec.~\ref{sec:gateaux} is independent of how the Gateaux derivatives in Eq.~\eqref{eq:def_gateaux} are evaluated, one may in principle combine the same protocol with different derivative engines (e.g., AD or jet arithmetic).
We emphasize that this remark concerns only the evaluation of the required mixed directional derivatives:
To our knowledge, there is presently no established AD-based response-function framework that, by itself, provides a systematic route from real-time dynamics to fully frequency-resolved multivariable kernels $\bar{\chi}^{(n)}(\omega_1,\dots,\omega_n)$.
The central difficulty lies not in differentiation, but in organizing and extracting the mixed functional derivatives needed for multivariable frequency reconstruction.
Accordingly, the contributions of this work are twofold: (1) a general reconstruction protocol that maps infinitesimal field-functional derivatives to frequency-resolved nonlinear response functions (Sec.~\ref{sec:gateaux}); and (2) the TEOM framework as an explicit EOM-level derivative engine that evaluates these derivatives in the strict $\epsilon \to 0$ limit, while providing a closed hierarchy whose term-by-term structure enables physically meaningful decompositions (e.g., injection versus shift currents) and equation-level numerical stabilization.

For the fully multivariable $n$th-order response targeted with $m=n-1$ independent perturbation fields, computing the complete set of mixed field derivatives requires propagating $\mathcal{O}(2^m)$ derivative components, and thus any backend that explicitly evaluates all mixed derivatives inherits at least the same exponential scaling in $m$.
The practical difference is therefore largely in representation and overhead:
An AD- or jet-based implementation could in principle use existing time-propagation codes by replacing scalar arithmetic with extended types, at the cost of increased arithmetic and memory overhead, whereas the TEOM implementation retains standard floating-point operations but evolves a larger coupled hierarchy of equations.
A detailed performance comparison is solver- and hardware-dependent and is beyond the scope of this work.

When one restricts to a single perturbation direction (i.e., a single field component with identical perturbation fields), the task reduces to a univariate $m$th derivative with respect to $\epsilon$, and both TEOM and generic AD derivative engines can admit jet implementations \cite{Berz1989} with polynomial (rather than exponential) scaling in $m$, as demonstrated for TEOM in Sec.~\ref{sec:duffing}.
Our main formalism in Sec.~\ref{sec:formalism} does not rely on this restriction, so it remains applicable to general frequency configurations and mixed derivatives.

Finally, we outline how the TEOM framework can be extended to nonequilibrium response problems in pump-driven transient states and periodically driven Floquet states \cite{Eckstein2008, DeFilippis2012, Lenarcic2014, Shao2016, Shinjo2018, Ono2018, Ono2019e, Kumar2019, Eskandari-asl2024, Ejima2022, Udono2023, Tohyama2023, Takubo2024, Shinjo2024, Shinjo2025, Volkov2023, Kim2024, Osterkorn2025}.
In such settings, a strong pump field may be included nonperturbatively in the underlying real-time EOM to generate a time-dependent reference trajectory, and the TEOM can then be constructed by differentiating the corresponding EOM with respect to an additional weak probe field.
The resulting response functions are no longer stationary equilibrium kernels:
In general, they retain an explicit dependence on the observation time, pump--probe delay, or Floquet phase.
This also changes the frequency-domain reconstruction, because the multivariable response function $\chi^{(n)}(t;t_1,\dots,t_n)$ cannot be reduced simply to a time-translation-invariant kernel.
A useful feature of the present time-domain formulation is that the temporally localized probe field $f(t)$ defines the probe time and delay explicitly before any Fourier or Wigner transformation is performed, as discussed for transient linear responses in Ref.\ \cite{Shao2016}.
This may provide a framework for formulating nonlinear response functions in nonequilibrium states.
A detailed development of this direction is beyond the scope of the present work and will be reported elsewhere.

\section{Summary} \label{sec:summary}
We have developed an EOM-level infinitesimal-variation framework that enables systematic, accurate, and efficient extraction of nonlinear response functions directly from real-time dynamics, without explicit evaluation of multipoint correlation functions or numerically unstable subtraction procedures.
The key step is to introduce the Gateaux derivative with respect to the perturbation field in function space, which yields a closed hierarchy of TEOM for mixed derivatives of the evolving state.
By integrating this hierarchy and applying tailored perturbation fields, multivariable frequency-resolved response kernels can be reconstructed beyond diagonal (harmonic) slices, while retaining a transparent structure that supports term-by-term physical interpretation and equation-level numerical stabilization.

Operationally, isolating the $n$th-order response requires $n-1$ directional derivatives in the weak-source formulation, whereas it requires $n$ directional derivatives in the zero-field source-derivative formulation.
The resulting TEOM hierarchy involves an exponential number of auxiliary dynamical variables in the fully general multivariable setting, but crucially avoids the factorial complexity inherent to diagrammatic or explicit-correlator formulations.
Moreover, in diagonal configurations where all perturbation directions are identical, the hierarchy collapses to a polynomially sized system, enabling access to very high response orders.

We have validated the method in representative quantum and classical settings, including benchmark comparisons for second- and third-order optical responses, frequency-resolved fifth-order response maps in an extended solid-state electron model, and high-order harmonic response functions up to the 49th order in a Duffing-type oscillator with controlled accuracy.
These results establish TEOM as a computationally efficient framework that interfaces naturally with state-of-the-art real-time solvers, making high-order and frequency-resolved nonlinear response functions practically accessible.
Beyond these examples, the TEOM framework demonstrates broad applicability across a wide range of quantum and classical dynamics in physics and related disciplines.

\begin{acknowledgments}
This work was supported by the Japan Society for the Promotion of Science (JSPS) KAKENHI Grants No.\ JP23K13052, No.\ JP24K00563, No.\ JP26K06993, and No.\ JP26K00646.
The numerical calculations were performed using the facilities of the Supercomputer Center, the Institute for Solid State Physics, the University of Tokyo.
\end{acknowledgments}

\section*{Data Availability}
A Julia implementation of the TEOM calculations for the Rice--Mele model discussed in Sec.~\ref{sec:RM} is publicly available on Zenodo \cite{ZenodoTEOMRiceMele}.
The repository contains scripts for real-time TEOM propagation and Fourier reconstruction of nonlinear current-response functions.

\appendix
\section{Window functions} \label{sec:window}
The window function $w_{\tau}(t)$ discussed in Sec.~\ref{sec:teom} is introduced for time gating (finite-time observation) and is chosen such that $w_{\tau}(t) \to 1$ as $\tau \to \infty$.
For the perturbation fields in Eqs.~\eqref{eq:gcos_t} and \eqref{eq:gsin_t}, we require the envelope $u_{\tau}(t)$ to satisfy $u_{\tau}(\omega) \to \delta(\omega)$ as $\tau \to \infty$ under the Fourier-transform convention in Eq.~\eqref{eq:u_to_delta}.
If we take the envelope to share the same shape as the window by setting
\begin{align}
u_{\tau}(t) = \frac{1}{2\pi} w_{\tau}(t),
\end{align}
then $u_{\tau}(t) \to 1/(2\pi)$ and hence $u_{\tau}(\omega) \to \delta(\omega)$ with unit weight.
We present several examples of $w_{\tau}(t)$ and their Fourier transforms below.
\begin{enumerate}
\item Gaussian window:
\begingroup\allowdisplaybreaks\begin{gather}
w_{\tau}(t) = \exp\left(-\frac{t^2}{2\tau^2}\right), \\
w_{\tau}(\omega) = \sqrt{2\pi}\, \tau \exp\left(-\frac{\omega^2 \tau^2}{2}\right).
\end{gather}\endgroup
\item Exponential window:
\begingroup\allowdisplaybreaks\begin{gather}
w_{\tau}(t) = \exp\left(-\frac{\vert t \vert}{\tau} \right), \\
w_{\tau}(\omega) = \frac{2\tau}{1+(\omega \tau)^2} = \frac{2 \tau^{-1}}{\omega^2 + \tau^{-2}}. \label{eq:window_exp_w}
\end{gather}\endgroup
\item Hann window:
\begingroup\allowdisplaybreaks\begin{gather}
w_{\tau}(t) = \varTheta(2\tau - \vert t \vert) \cos^2 \left(\frac{\pi t}{4\tau}\right), \\
w_{\tau}(\omega) = \frac{\pi^2 \tau \sin(2\omega \tau)}{\pi^2 \omega \tau - 4 (\omega \tau)^3}.
\end{gather}\endgroup
\end{enumerate}
Here, $\varTheta$ is the unit step function.
The Gaussian window has the desirable property of decaying rapidly in both the time and frequency domains.
The exponential window decays more slowly, but its compatibility with the RTA makes it a suitable option when one is interested in comparisons with frequency-domain methods and in the details of dissipation and relaxation processes.
The Hann window has finite support in the time domain, which is advantageous in time-evolution calculations, while its Fourier transform decays as $(\omega \tau)^{-3}$, faster than that of the exponential window but slower than that of the Gaussian window.

\section{Continuous-wave driving method for extracting harmonic response functions} \label{sec:cw}
Here, we briefly present a procedure for obtaining harmonic response functions from the time evolution in the presence of a cw external field.

Equation~\eqref{eq:Qn_w} can be written using the symmetrized response function as
\begin{align}
&\langle Q_{\alpha} \rangle^{(n)}(\omega) \notag \\
&= \frac{1}{n!} \sum_{\{\beta_i\!\}} \int_{-\infty}^{\infty} \frac{\mathrm{d}\omega_1 \cdots \mathrm{d}\omega_n}{(2\pi)^{n-1}} \delta(\omega_1 + \cdots + \omega_n - \omega) \notag \\
&\quad \times \bar{\chi}_{\alpha \beta_1 \cdots \beta_n\!}^{(n)}(\omega_1, \dots, \omega_n) f_{\beta_1\!}(\omega_1) \cdots f_{\beta_n\!}(\omega_n),
\label{eq:Qn_w_sym}
\end{align}
where the prefactor $1/n!$ compensates for the overcounting that arises from the $n! = |\mathfrak{S}_n|$ elements of the symmetric group.
In what follows, we omit the indices $\alpha$ and $\beta$ for simplicity and focus on a single component of the physical quantity and the external field.

We introduce a cw external field with frequency $\omega_{\mathrm{cw}}$,
\begin{gather}
f(t) = F_0 \cos(\omega_{\mathrm{cw}}t - \phi_{\mathrm{cw}}), \\
f(\omega) = 2\pi \frac{F_0}{2} \left[ \mathrm{e}^{\mathrm{i}\phi_{\mathrm{cw}}} \delta(\omega - \omega_{\mathrm{cw}}) + \mathrm{e}^{-\mathrm{i}\phi_{\mathrm{cw}}} \delta(\omega + \omega_{\mathrm{cw}}) \right],
\end{gather}
where $F_0$ and $\phi_{\mathrm{cw}}$ denote the amplitude and phase of the field, respectively.
By substituting $f(\omega)$ into Eq.~\eqref{eq:Qn_w_sym} and collecting the contribution at frequency $+ n \omega_{\mathrm{cw}}$, we obtain
\begin{align}
\langle Q \rangle^{(n)}(\omega)
&= \frac{2\pi}{n!} \left(\frac{F_0}{2}\right)^{n} \delta(\omega - n\omega_{\mathrm{cw}}) \bar{\chi}_{\mathrm{cw}}^{(n)} \mathrm{e}^{\mathrm{i}n\phi_{\mathrm{cw}}} \notag \\
&\quad + \text{(terms at other frequencies)},
\label{eq:Qn_w_cw}
\end{align}
where $\bar{\chi}_{\mathrm{cw}}^{(n)}$ is a shorthand for $\bar{\chi}^{(n)}(\omega_{\mathrm{cw}}, \dots, \omega_{\mathrm{cw}})$.
This indicates that the $n$th-order harmonic response, i.e., the oscillatory response at frequency $n\omega_{\mathrm{cw}}$, can be expressed in the time domain as
\begin{align}
\langle Q \rangle^{(n)}(t)
= \frac{\vert \bar{\chi}_{\mathrm{cw}}^{(n)} \vert}{n!} \frac{F_0^n}{2^{n-1}} \cos\left[n(\omega_{\mathrm{cw}}t - \phi_{\mathrm{cw}}) - \arg \bar{\chi}_{\mathrm{cw}}^{(n)}\right].
\label{eq:Qn_t_cw}
\end{align}
To derive Eq.~\eqref{eq:Qn_t_cw} from Eq.~\eqref{eq:Qn_w_cw}, we used Eq.~\eqref{eq:chi_conjugate}.

When the RTA is introduced, the system driven by the cw field eventually reaches a steady state.
In the steady state, the time-domain response can be expressed as
\begin{align}
\langle Q \rangle(t) = \sum_{n\geq 0} A_n \cos(n\omega_{\mathrm{cw}}t - \phi_n),
\end{align}
where the amplitude $A_n$ ($\geq 0$) and phase $\phi_n$ of the $n$th harmonic can be extracted from a discrete Fourier transform of $\langle Q \rangle(t)$ (see Refs.\ \cite{Ono2024, Ono2025b} for details).
Therefore, the $n$th-order response function is obtained as
\begin{align}
\bar{\chi}^{(n)}(\omega_{\mathrm{cw}}, \dots, \omega_{\mathrm{cw}})
= \frac{2^{n-1} n! A_n}{F_0^n} \mathrm{e}^{\mathrm{i}(\phi_n - n \phi_{\mathrm{cw}})}.
\end{align}
In practical calculations, one should first examine the dependence of $A_n$ on the field amplitude $F_0$ and then choose a sufficiently small $F_0$ such that the response lies in the perturbative regime; namely, the relation $A_n \propto F_0^n$ is satisfied.

\section{Incident polarization-angle dependence of HHG intensity in the perturbative regime} \label{sec:In}
We first consider the response of the current density $\langle Q_\alpha \rangle = \langle j_\alpha \rangle$ to the vector potential $f_{\alpha} = A_{\alpha}$ in a two-dimensional system ($\alpha \in \{x,y\}$).
We then derive the expressions for the intensity of HHG as functions of the polarization angle of the incident light, by imposing sixfold symmetry on the response functions.

The vector potential of the incident light is taken to be
\begin{gather}
\bm{A}(t) = \bm{F}_0 \cos(\omega_{\mathrm{cw}}t), \\
\bm{F}_0 = (F_x, F_y) = (F_0 \cos\psi, F_0 \sin\psi),
\end{gather}
where $F_0$ and $\psi$ denote the amplitude and the polarization angle, respectively.
By restoring the indices omitted in Appendix~\ref{sec:cw} and extracting the contribution at frequency $+n\omega_{\mathrm{cw}}$, we obtain
\begin{align}
\langle j_\alpha \rangle^{(n)}(\omega)
&= \frac{2\pi}{2^{n} n!} \sum_{\{\alpha_i\}} F_{\alpha_1} \cdots F_{\alpha_n} \bar{\chi}_{\alpha\alpha_1 \cdots \alpha_n}^{(n)} \delta(\omega - n\omega_{\mathrm{cw}}) \notag \\
&\quad + \text{(terms at other frequencies)},
\end{align}
where $\bar{\chi}_{\alpha \alpha_1 \cdots \alpha_n\!}^{(n)}$ represents $\bar{\chi}_{\alpha \alpha_1 \cdots \alpha_n\!}^{(n)}(\omega_{\mathrm{cw}}, \dots, \omega_{\mathrm{cw}})$.
In the time domain, the current density of the $n$th harmonic is then given by
\begin{align}
\langle j_\alpha \rangle^{(n)}(t)
&= \frac{1}{2^{n-1} n!} \sum_{\{\alpha_i\}} F_{\alpha_1} \cdots F_{\alpha_n} \notag \\
&\quad \times \vert\bar{\chi}_{\alpha\alpha_1 \cdots \alpha_n\!}^{(n)}\vert \cos(n\omega_{\mathrm{cw}}t - \phi_{\alpha\alpha_1 \cdots \alpha_n} ),
\end{align}
where $\phi_{\alpha\alpha_1 \cdots \alpha_n\!} = \arg \bar{\chi}_{\alpha\alpha_1 \cdots \alpha_n}^{(n)}$.
The intensity $I_n$ of the $n$th harmonic is proportional to the time average (denoted by $\overline{\bullet}$) of the squared time derivative of the current density, namely,
\begin{align}
I_n &= \overline{\Vert \partial \langle \bm{j} \rangle^{(n)}(t)/\partial t \Vert^2} \notag \\
&= \frac{(n\omega_{\mathrm{cw}})^2}{2^{2n-2} n!^2} \sum_{\alpha} \sum_{\{\alpha_i, \beta_i\}} F_{\alpha_1} \cdots F_{\alpha_n} F_{\beta_1} \cdots F_{\beta_n} \notag \\
&\quad \times \vert\bar{\chi}_{\alpha\alpha_1 \cdots \alpha_n\!}^{(n)} \vert \vert \bar{\chi}_{\alpha\beta_1 \cdots \beta_n\!}^{(n)}\vert \notag \\
&\quad \times \overline{\sin(n\omega_{\mathrm{cw}}t - \phi_{\alpha\alpha_1 \cdots \alpha_n} ) \sin(n\omega_{\mathrm{cw}}t - \phi_{\alpha\beta_1 \cdots \beta_n} )} \notag \\
&= \frac{(n\omega_{\mathrm{cw}})^2}{2^{2n-1} n!^2} \sum_{\alpha} \sum_{\{\alpha_i, \beta_i\}} F_{\alpha_1} \cdots F_{\alpha_n} F_{\beta_1} \cdots F_{\beta_n} \notag \\
&\quad \times \re(\bar{\chi}_{\alpha\alpha_1 \cdots \alpha_n\!}^{(n)*} \bar{\chi}_{\alpha\beta_1 \cdots \beta_n\!}^{(n)}).
\end{align}

The $n$th-order optical response function $\chi^{(n)}$ is a polar tensor of rank $n+1$, and the number of its independent components is reduced by crystal symmetry.
Let $M$ denote the orthogonal matrix representing a symmetry operation.
Then the response function satisfies
\begin{align}
\chi_{\alpha_0\alpha_1\cdots\alpha_n\!}^{(n)}
= \sum_{\{\beta_i\!\}} M_{\alpha_0\beta_0} M_{\alpha_1\beta_1} \cdots M_{\alpha_n\beta_n} \chi_{\beta_0\beta_1\cdots\beta_n\!}^{(n)}.
\label{eq:tensor_symmetry}
\end{align}

In Sec.~\ref{sec:higher-order}, we considered a four-band model that preserves sixfold rotational symmetry.
By applying Eq.~\eqref{eq:tensor_symmetry} to a rotation by $\pi/3$ about the $z$ axis, we derive the polarization-angle dependence of the harmonic intensity for this model.
For the first, third, and fifth harmonics, the results are given by
\begin{align}
I_1 &= \frac{\omega_{\mathrm{cw}}^2}{2} \bigl(\vert\chi_{xx}^{(1)}\vert^2 + \vert\chi_{yx}^{(1)}\vert^2\bigr) F_0^2, \label{eq:I1} \\
I_3 &= \frac{(3\omega_{\mathrm{cw}})^2}{2^5 3!^2} \bigl(\vert\bar{\chi}_{xxxx}^{(3)}\vert^2 + \vert\bar{\chi}_{yxxx}^{(3)}\vert^2\bigr) F_0^6, \label{eq:I3} \\
I_5 &= \frac{(5\omega_{\mathrm{cw}})^2}{2^9 5!^2} \bigl[C_0 + C_+ \cos(6\psi) + C_- \sin(6\psi)\bigr] F_0^{10}, \label{eq:I5}
\end{align}
respectively.
The coefficients in Eq.~\eqref{eq:I5} are defined by
\begin{align}
C_0 &= \frac{1}{8} \bigl[
13 \vert \bar{\chi}_{xxxxxx}^{(5)} \vert^2
+ 52 \vert \bar{\chi}_{xxxxxy}^{(5)} \vert^2
+ 52 \vert \bar{\chi}_{xxxyyy}^{(5)} \vert^2 \notag \\
&\quad
+ 25 \vert \bar{\chi}_{xxyyyy}^{(5)} \vert^2
- 30 \re(\bar{\chi}_{xxyyyy}^{(5)*} \bar{\chi}_{xxxxxx}^{(5)}) \notag \\
&\quad
+ 96 \re(\bar{\chi}_{xxxyyy}^{(5)*} \bar{\chi}_{xxxxxy}^{(5)})
\bigr]
, \label{eq:I5_C0} \\
C_+ &= -\frac{5}{8} \bigl[
\vert \bar{\chi}_{xxxxxx}^{(5)} \vert^2
+ 4 \vert \bar{\chi}_{xxxxxy}^{(5)} \vert^2
- 4 \vert \bar{\chi}_{xxxyyy}^{(5)} \vert^2 \notag \\
&\quad
+ 5 \vert \bar{\chi}_{xxyyyy}^{(5)} \vert^2
- 6 \re(\bar{\chi}_{xxyyyy}^{(5)*} \bar{\chi}_{xxxxxx}^{(5)})
\bigr], \label{eq:I5_Cp} \\
C_- &= -\frac{5}{2} \bigl[
\re(\bar{\chi}_{xxxyyy}^{(5)*} \bar{\chi}_{xxxxxx}^{(5)})
- 2 \re(\bar{\chi}_{xxyyyy}^{(5)*} \bar{\chi}_{xxxxxy}^{(5)}) \notag \\
&\quad
- 3 \re(\bar{\chi}_{xxyyyy}^{(5)*} \bar{\chi}_{xxxyyy}^{(5)})
\bigr]. \label{eq:I5_Cm}
\end{align}
Using a constraint from the sixfold symmetry,
\begin{align}
\bar{\chi}_{yxxxxx}^{(5)}
= -2 \bar{\chi}_{xxxxxy}^{(5)} - 3 \bar{\chi}_{xxxyyy}^{(5)},
\end{align}
one can rewrite the expression for $I_5$ at $\psi = 0$ as
\begin{align}
I_5\vert_{\psi = 0} &= \frac{(5\omega_{\mathrm{cw}})^2}{2^9 5!^2} \bigl(\vert\bar{\chi}_{xxxxxx}^{(5)}\vert^2 + \vert\bar{\chi}_{yxxxxx}^{(5)}\vert^2\bigr) F_0^{10}. \label{eq:I5_psi0}
\end{align}
Note that the expressions for $I_1$ and $I_3$ are identical to those presented in Ref.\ \cite{Ono2024}, whereas Eqs.~\eqref{eq:I5_C0}--\eqref{eq:I5_Cm} for $I_5$ are corrected here by fully incorporating interference terms that were previously overlooked.

\section{Green's-function evaluation of the nonlinear response of the Duffing oscillator} \label{sec:duffing_green}
We summarize a frequency-domain Green's-function approach to the nonlinear response functions of the Duffing oscillator.
Throughout, we adopt the Fourier-transform convention
\begin{align}
f(\omega) = \int_{-\infty}^{\infty} \mathrm{d}t\, \mathrm{e}^{\mathrm{i} \omega t} f(t), \quad
f(t) = \int_{-\infty}^{\infty} \frac{\mathrm{d}\omega}{2\pi}\, \mathrm{e}^{-\mathrm{i}\omega t} f(\omega),
\end{align}
consistent with the main text.

We consider the scalar Duffing oscillator driven by an external force $f(t)$, defined in Eq.~\eqref{eq:duffing}.
Introducing the linear operator
\begin{align}
L_t \coloneq \frac{\mathrm{d}^2}{\mathrm{d}t^2} + \gamma \frac{\mathrm{d}}{\mathrm{d}t} + \omega_0^2,
\end{align}
we can recast the EOM [Eq.~\eqref{eq:duffing}] as
\begin{align}
L_t x(t) = f(t) - \alpha x(t)^3.
\label{eq:duffing_operator_form_t}
\end{align}
The corresponding retarded Green's function $G(t)$ is defined by
\begin{align}
L_t G(t) = \delta(t)
\end{align}
with $G(t < 0) = 0$.
In the frequency domain, one obtains
\begin{align}
G(\omega) = \frac{1}{(\omega_0^2-\omega^2) - \mathrm{i}\gamma\omega}.
\label{eq:duffing_Gw}
\end{align}
Equation~\eqref{eq:duffing_operator_form_t} then yields the formal solution
\begin{align}
x(\omega) = G(\omega)\bigl[ f(\omega)-\alpha (x*x*x)(\omega) \bigr],
\label{eq:duffing_integral_form_w}
\end{align}
where $(x*x*x)(\omega)$ denotes the triple convolution in frequency,
\begin{align}
&(x*x*x)(\omega) \notag \\
&= \int_{-\infty}^{\infty} \frac{\mathrm{d}\omega_1 \mathrm{d}\omega_2 \mathrm{d}\omega_3}{(2\pi)^2}
\delta(\omega_1 + \omega_2 + \omega_3 - \omega) x(\omega_1) x(\omega_2) x(\omega_3).
\end{align}

In a perturbative regime, we decompose the response into contributions that are homogeneous in the drive amplitude,
\begin{align}
x(\omega) = x^{(0)}(\omega) + \sum_{n=1}^{\infty} x^{(n)}(\omega),
\quad x^{(n)} = \mathcal{O}(f^n),
\end{align}
where $x^{(0)}$ is the solution for $f = 0$ (here, $x^{(0)} = 0$).
For $n \geq 1$, the $n$th-order contribution can be written in the standard Volterra-kernel form
\begin{align}
x^{(n)}(\omega)
&= \int \frac{\mathrm{d}\omega_1 \cdots \mathrm{d}\omega_n}{(2\pi)^{n-1}}
\delta(\omega_1 + \cdots + \omega_n - \omega) \notag \\
&\quad \times \chi^{(n)}(\omega_1,\dots,\omega_n) f(\omega_1) \cdots f(\omega_n).
\label{eq:def_chi_duffing}
\end{align}
In the main text, we work with the fully symmetrized kernel [Eq.~\eqref{eq:symmetrized_chi}] defined as the sum over all permutations without the $1/n!$ normalization.
In the present scalar problem, Eq.~\eqref{eq:def_chi_duffing} is invariant under permutations of the integration variables $\omega_i \leftrightarrow \omega_j$.
Therefore, only the fully symmetric part of $\chi^{(n)}$ contributes, and we can, without loss of generality, take $\chi^{(n)}(\omega_1,\dots,\omega_n)$ to be permutation symmetric.
Under this choice, our convention implies $\bar{\chi}^{(n)} = n! \chi^{(n)}$.
We nevertheless keep the notation $\bar{\chi}^{(n)}$ because it matches the convention used in the numerical extraction in the main text,
and because the recursion below naturally produces $\bar{\chi}^{(n)}$ as a sum over labeled (ordered) tree diagrams.

At linear order, Eq.~\eqref{eq:duffing_integral_form_w} gives
\begin{align}
x^{(1)}(\omega) = G(\omega)f(\omega), \quad \chi^{(1)}(\omega) = G(\omega).
\label{eq:chi1_duffing}
\end{align}

To third order, one substitutes $x \to x^{(1)}$ into the cubic term of Eq.~\eqref{eq:duffing_integral_form_w}:
\begin{align}
x^{(3)}(\omega) = -\alpha G(\omega) (x^{(1)}*x^{(1)}*x^{(1)})(\omega).
\end{align}
Using Eq.~\eqref{eq:chi1_duffing} and comparing with Eq.~\eqref{eq:def_chi_duffing}, one finds
\begin{align}
&\chi^{(3)}(\omega_1,\omega_2,\omega_3) \notag \\
&= -\alpha G(\omega_1+\omega_2+\omega_3) G(\omega_1) G(\omega_2) G(\omega_3),
\label{eq:chi3_duffing}
\end{align}
and hence
\begin{align}
&\bar{\chi}^{(3)}(\omega_1,\omega_2,\omega_3) \notag \\
&= -3! \alpha G(\omega_1+\omega_2+\omega_3) G(\omega_1) G(\omega_2) G(\omega_3).
\label{eq:chibar3_duffing}
\end{align}
The factorial originates purely from the convention in Eq.~\eqref{eq:symmetrized_chi} (sum over permutations).

At higher odd orders, the cubic nonlinearity generates nested contributions built from lower-order responses.
This is already visible at fifth order.
Writing $x = x^{(1)} + x^{(3)} + x^{(5)} + \cdots$ and collecting terms of order $f^5$ in Eq.~\eqref{eq:duffing_integral_form_w}, we obtain
\begin{align}
x^{(5)}(\omega)
= -3\alpha G(\omega) \bigl(x^{(1)}*x^{(1)}*x^{(3)}\bigr)(\omega).
\label{eq:x5_nested}
\end{align}
The prefactor $3$ arises from the algebraic expansion $(x^{(1)}+x^{(3)}+\cdots)^3 = (x^{(1)})^3 + 3 (x^{(1)})^2 x^{(3)} + \cdots$.
Upon substituting Eq.~\eqref{eq:chi3_duffing} for $x^{(3)}$, one finds that $x^{(5)}$ involves an inner third-order response constructed from a chosen triple of input frequencies, while the remaining two frequencies attach directly to the outer cubic vertex.
Thus, even at $n=5$, the kernel $\chi^{(5)}(\omega_1,\dots,\omega_5)$ is a sum over all ways of selecting and grouping subsets of the input frequencies, corresponding to distinct rooted ternary-tree contractions.
This subset bookkeeping becomes rapidly more cumbersome at higher orders, motivating the subset recursion introduced below.

Let the $n$ input frequencies be $\{\omega_i\}_{i=1}^n$ and define the partial sum
\begin{align}
\omega_S \equiv \sum_{i \in S} \omega_i
\end{align}
for any nonempty subset $S \subseteq \{1, \dots, n\}$.
We define a subset-resolved quantity $\varPhi(S)$ by
\begin{align}
\varPhi(S) \equiv \bar{\chi}^{(|S|)}(\{\omega_i\}_{i\in S}),
\label{eq:Phi_def}
\end{align}
i.e., the fully symmetrized response kernel associated with the multiset of frequencies in $S$.
Then the cubic nonlinearity implies the recursion
\begin{align}
\varPhi(S) = G(\omega_S)
\end{align}
for $|S|=1$, and
\begin{align}
\varPhi(S) = -\alpha G(\omega_S) \sum_{\substack{A,B,C\neq\varnothing\\ A\sqcup B\sqcup C = S}} \varPhi(A) \varPhi(B) \varPhi(C)
\label{eq:Phi_recursion}
\end{align}
for $|S| \geq 2$.
Here, the sum runs over ordered three-partitions $(A,B,C)$ of $S$ into nonempty disjoint subsets.
We expand around the unforced equilibrium $x^{(0)} = 0$, so partitions with an empty block do not contribute and we restrict to $A, B, C \neq \varnothing$.
This ordered-partition convention is precisely what produces the unnormalized full symmetrization in Eq.~\eqref{eq:symmetrized_chi}.
In particular, $\varPhi(S)=0$ for even $|S|$.
Indeed, $\varPhi(S)=0$ for $|S| = 2$ because there is no ordered three-partition of a two-element set into three nonempty blocks.
Assuming $\varPhi$ vanishes for all even subset sizes smaller than $|S|$, any ordered three-partition of an even-sized set $S$ must contain at least one even-sized block, whose $\varPhi$ is zero by the induction hypothesis; hence, the sum in Eq.~\eqref{eq:Phi_recursion} vanishes.

For the full $n$th-order kernel, we take $S = \{1,\dots,n\}$:
\begin{align}
\bar{\chi}^{(n)}(\omega_1,\dots,\omega_n) = \varPhi(\{1,\dots,n\}).
\label{eq:chibar_full_from_Phi}
\end{align}
Equation~\eqref{eq:Phi_recursion} is equivalent to a sum over all rooted ternary trees with $n$ labeled leaves.
Each cubic vertex contributes a factor $(-\alpha)$, and each internal line (including the root) contributes a propagator $G(\omega_S)$ evaluated at the sum of frequencies flowing through that line.

For example, at fifth order ($n=5$) the recursion reduces to partitions of the form $(1, 1, 3)$ (and permutations thereof) at the top vertex, where the size-$3$ block is evaluated using Eq.~\eqref{eq:chibar3_duffing}.
This illustrates how Eq.~\eqref{eq:Phi_recursion} systematically generates the correct combinatorial weights at $n \geq 5$ without explicitly enumerating diagrams by hand.

\bibliography{reference}

\end{document}